\documentclass{aa}                   %  A & A  Latex Macro V10
\usepackage{graphicx}                %  Include graphics
\usepackage{txfonts}                 %  Postscript Fonts
\newcommand{\Halpha}{H$\alpha$}
\newcommand{\Hbeta}{H$\beta$}
\newcommand{\Hgamma}{H$\gamma$}
\newcommand{\kms}{km\,s$^{-1}$}
\newcommand{\ms}{m\,s$^{-1}$}

\begin{document}

\title{Stellar rotation, binarity, and lithium\\ in the open cluster IC\,4756\thanks{The CoRoT space mission has
been developed and is operated by CNES, with the contribution of
Austria, Belgium, Brazil, ESA (RSSD and Science Programme), Germany
and Spain. Partly based on data obtained with the STELLA robotic
observatory in Tenerife, an AIP facility jointly operated by AIP and
IAC.}}

\author{K. G.~Strassmeier\inst{1}, J.~Weingrill\inst{1}, T.~Granzer\inst{1}, G. Bihain\inst{1}, M. Weber\inst{1}, \and S. A.~Barnes\inst{1,2}}

\offprints{K. G. Strassmeier}

\institute{Leibniz-Institute for Astrophysics Potsdam (AIP), An der
Sternwarte 16, D-14482 Potsdam, Germany, \and
Space Science Institute, 4750 Walnut Street \#205, Boulder, CO 80301, U.S.A.\\
\email{KStrassmeier, JWeingrill, TGranzer, GBihain, MWeber, SBarnes@aip.de}}

\date{Received ... ; accepted ...}

\abstract{An important aspect in the evolutionary scenario of cool stars is their rotation and the rotationally induced magnetic activity and interior mixing. Stars in open clusters  are particularly useful tracers for these aspects because of their known ages.}{We aim to characterize the open cluster IC\,4756 and measure stellar rotation periods and surface differential rotation for a sample of its member stars.} {Thirty-seven cluster stars were observed continuously with the CoRoT satellite for 78 days in 2010. Follow-up high-resolution spectroscopy of the CoRoT targets and deep Str\"omgren $uvby\beta$ and \Halpha\ photometry of the entire cluster were obtained with our robotic STELLA facility and its echelle spectrograph and wide-field imager, respectively.}{We determined high-precision photometric periods for 27 of the 37 CoRoT targets and found values between 0.155 and 11.4\,days. Twenty of these are rotation periods. Twelve targets are spectroscopic binaries of which 11 were previously unknown; orbits are given for six of them. Six targets were found that show evidence of differential rotation with $\Delta\Omega/\Omega$ in the range 0.04--0.15. Five stars are non-radially pulsating stars with fundamental periods of below 1\,d, two stars are semi-contact binaries, and one target is a micro-flaring star that also shows rotational modulation.  Nine stars in total were not considered members because of much redder color(s) and deviant radial velocities with respect to the cluster mean. \Halpha\ photometry indicates that the cluster ensemble does not contain magnetically over-active stars. The cluster average metallicity is --0.08$\pm$0.06 (rms) and its logarithmic lithium abundance for 12 G-dwarf stars is 2.39$\pm$0.17 (rms).}{The cluster is 890$\pm$70\,Myrs old with an average turn-off mass of 1.8\,M$_\odot$ and a solar or slightly subsolar metallicity. The distance modulus is 8\fm02 and the average reddening $E(b-y)=0\fm16$. The cluster is masked by a very inhomogeneous foreground and background dust distribution and our survey covered 38\%\ of the cluster. The average lithium abundance and the rotation periods in the range 1.7 to 11.4\,d are consistent with the cluster age.}

\keywords{stars: rotation -- stars: activity of -- stars: late-type -- starspots -- stars:oscillations -- open clusters: IC~4756 }

\authorrunning{Strassmeier et al.}
\titlerunning{Stellar rotation in the open cluster IC\,4756}

\maketitle
%------------------------------------------------------------------------

\section{Introduction}

Stellar surface rotation and differential rotation are crucial to our understanding of magnetic activity of late-type stars. The rotation periods of stars in open clusters with well-determined ages will not only confirm  the gyro-ages put forward in previous years by Barnes (\cite{bar07}, \cite{bar10}), but will also help to quantitatively establish the angular-momentum transport mechanism as a function of stellar evolution (Barnes \& Kim \cite{bar:kim}, Gallet \& Bouvier \cite{gal:bou}, Meibom et al. \cite{mei:cs17}, \cite{mei:nature}). The magnitude and sign of the surface differential rotation pattern can even specify the role of the stellar magnetic field because differential rotation constrains the large-scale magnetic surface configurations, and will tell us about the underlying dynamo process (e.g., K\"uker et al. \cite{kuk:rud}, K\"apyl\"a et al.~\cite{kap:man}, Hathaway et al. \cite{hath}, R\"udiger et al. \cite{rue:kue}).

Rotation can be recovered from the spot modulation in time-series photometry, either from the ground via  decade-long coverage with small robotic telescopes (e.g., Henry \cite{hen}, Ol\'ah et al. \cite{olah}, Strassmeier \cite{aarev}) or from space with comparatively short runs but basically perfect sampling (e.g., for CoRoT-2a; Fr\"ohlich et al. \cite{froh}). For space-based observations, we are confident that even if a spot configuration is less favorable than CoRoT-2a, a unique interpretation is possible from the clean window function and the high photometric precision (e.g., Reinhold et al. \cite{rein}). Furthermore, inversion of a time series using a spot-modeling approach also allows the extraction of spot longitudes and fractional spot coverage (e.g., Strassmeier \& Bopp \cite{str:bop}, Lanza et al. \cite{lanza}, Savanov \& Dmitrienko \cite{sav:dim}, Roettenbacher et al. \cite{roett}).

An essential role for rotational studies of open clusters is to relate stellar properties to the general cluster properties. Among these properties are a cluster's turn-off age and average metallicity and also the lithium abundance and dispersion of its G dwarfs in order to relate eventual models to the Sun as a star. While space-based photometry is clearly superior for periodicities of various sorts, ground-based multi-band photometry and follow-up spectroscopy are essential in determining membership, abundances, and binarity (see, e.g., Meibom et al. \cite{mei:mat}, Milliman et al. \cite{mil:mat}). We have therefore used the CoRoT satellite (Baglin et al. \cite{bag}) in its sixth long run to observe selected and bona fide member stars of the open cluster IC\,4756 that happened to be within the exoplanet field of view of the satellite. From the ground, we employed our robotic STELLA facility with its wide-field imager and echelle spectrograph. With an age of $\approx$800~Myrs (Salaris et al. \cite{sal:wei}) IC\,4756 is nearly coeval with the Hyades where spot activity is weak but still present. The expected range of rotation periods for single stars is 2--20 days while the suggested turn-off mass is around 1.8--1.9~M$_\odot$ (Mermillod \& Mayor \cite{merm90}). Precise periods for IC\,4756 could be important for our understanding of rotational evolution of low-mass stars because all current rotational evolutionary models for above age are constrained (and restricted) by the statistics driven by the Hyades cluster (see Irwin \& Bouvier \cite{irw:bou}).

In the current paper, we present CoRoT time-series photometry for 37 of our initial 38 targets in the field of IC\,4756 - one of the initial 38 targets was not observed in the end - and determine precise rotational periods for 20 of them. Deep Str\"omgren $uvby\beta$ and H$\alpha$ photometry was obtained with our robotic STELLA facility to obtain a color-magnitude diagram and also to determine effective temperatures, gravities, metallicities, and an \Halpha -based activity measure for the detectable cluster members. STELLA has also been used to obtain high-resolution spectra to determine radial velocities, lithium abundances, and other astrophysical properties. One target, HSS\,348, has been observed and is being analyzed in more detail in a separate paper.

%------------------------------ Table 1:  COROT & STELLA sample
\begin{table*}[!tbh]
\begin{flushleft}
\caption{CoRoT sample of cluster stars, NOMAD positions, Herzog et al. (\cite{herz}) magnitudes, and STELLA/SES radial velocities. }\label{T1}
\begin{tabular}{llllllllllll}
\hline \noalign{\smallskip}
No. & HSS & USNO-B1 & Cluster & RA2000 & DEC2000 & $V$ & $B-V$ & $v_{\rm r}$ & $\sigma_{v_{\rm r}}$ & $N_{v_{\rm r}}$ & Notes$^c$\\
    &     &          &  member & (h:m:s)& ($^\circ$ : ' : '')& \multicolumn{2}{c}{(mag)} & \multicolumn{2}{c}{(\kms )} & & \\
\noalign{\smallskip}\hline \noalign{\smallskip}
 1 &  73 & 0956-0376800 & Y & 18:37:24.12 & 05:41:19.75 &  9.49 & 0.48 &  --25:   & 5  & 2 & \dots \\
 2$^a$ & 314 & 0953-0379344 & Y & 18:39:26.60 & 05:21:31.89 &  9.64 & 1.03 &  --24.6 & 0.06 & 39& SB1,giant\\
 3 & 138 & 0955-0379553 & Y & 18:38:00.55 & 05:35:19.68 &  9.88 & 0.45 &  --20: & 10 &  2&\dots \\
 4 & 106 & 0954-0378316 & Y & 18:37:40.39 & 05:29:44.92 & 10.04 & 0.44 &  --29.6 & 2.1 &  34& SB1\\
 5 & 108 & 0953-0377008 & Y & 18:37:41.29 & 05:19:11.35 & 11.01 & 0.40 &  --20.4 & 3.6  &  40 &SB1,Puls \\
 6 & 113 & 0955-0379309 & Y & 18:37:46.27 & 05:32:05.82 & 11.40 & 0.47 &  --11.2 & 1.5  & 9 & Puls\\
 7 & 112 & 0951-0373188 & Y  & 18:37:46.16 & 05:09:23.98 & 11.44 & 0.40 & --12.7 & 1.9  & 6 &\dots \\
 8 & 107 & 0953-0377007 & Y & 18:37:41.23 & 05:21:25.56 & 11.45 & 1.01 &  --25.8 & 0.4 &  52 & SB2,R,Li\\
 9 & 356 & 0954-0381320 & N & 18:39:47.36 & 05:28:33.02 & 11.48 & 1.46 &  --19.28 & 0.06 & 14 &Giant\\
10 & 241 & 0955-0380978 & Y & 18:38:47.87 & 05:32:29.62 & 11.63 & 0.56 &  --15.7 & 10  & 6 &SB,Puls\\
11 & 269 & 0954-0379835 & Y & 18:39:00.12 & 05:26:59.53 & 11.69 & 0.61 &  --23.6 & 3.8 & 9 & R,see\,App.\,B\\
12$^b$ & 208 & 0950-0377215 & N & 18:38:39.53 & 05:03:03.13 & 11.77 & 0.96 &  --1  & 12 & 5 &Puls\\
13 & 270 & 0950-0377687 & ? & 18:39:00.62 & 05:02:16.30 & 11.78 & 0.33 &  --22.2  & 2.4  & 4 &Puls\\
14 &  57 & 0956-0376593 & Y & 18:37:15.02 & 05:40:35.94 & 11.85 & 0.82 &  --22.1 & 0.9 & 6 &R \\
15$^b$ & 348 & 0950-0378805 & Y & 18:39:44.29 & 05:01:23.02 & 11.96 & 0.76 &  --23.8  & 20  & 18 & SB2,EB?,R\\
16 &  95 & 0957-0380623 & Y & 18:37:36.96 & 05:45:42.95 & 12.01 & 0.44 &  --22.8  & 2.3  & 4 & Li,R\\
17 & 174 & 0955-0379936 & Y & 18:38:19.73 & 05:30:03.89 & 12.30 & 0.40 &  --18.4 & 0.37  & 10 & SB2,R,Li\\
18 & 183 & 0956-0378250 & Y & 18:38:24.95 & 05:36:24.16 & 12.47 & 0.07 &  --24.5 & 0.6 & 6 &R\\
19 & 185 & 0956-0378356 & N & 18:38:27.13 & 05:36:03.35 & 12.58 & 0.83 &  --32.47  & 0.03  & 5 & Giant\\
20 & \dots & 0957-0380847 & N & 18:37:48.01 & 05:47:49.09 & 12.79 & 0.32 & +45.73   & 0.06  & 6 &Giant \\
21 & 194 & 0954-0379315 & Y & 18:38:32.23 & 05:29:25.22 & 12.84 & 0.35 &  --25.2 & 0.2 & 7 &R\\
22 & 158 & 0954-0378840 & N & 18:38:09.84 & 05:28:56.14 & 12.87 & 0.07 &  --1.5  & 14 & 5 & SB2,EB\\
23 &  89 & 0957-0380578 & N & 18:37:34.51 & 05:47:38.69 & 12.89 & 0.83 &  +34.4 & 0.1  & 5 & Giant\\
24$^b$ & 198 & 0949-0372664 & N & 18:38:35.04 & 04:57:28.15 & 13.00 & 1.01 & +39.63   & 0.14  & 5 &R,Giant\\
25 & 209 & 0953-0378197 & Y & 18:38:39.38 & 05:22:24.24 & 13.00 & 0.26 &  --23.9  & 0.6  & 5 &SB?,R\\
26 & 259 & 0954-0379730 & Y & 18:38:54.62 & 05:28:27.80 & 13.04 & 0.25 &  --24.37 & 0.13  & 6 &R,Li\\
27 & 284 & 0952-0377541 & Y & 18:39:07.40 & 05:13:46.96 & 13.14 & 0.20 &  --23.3 & 1.0 & 5 & SB?,R,Li\\
28 & 189 & 0954-0379240 & Y & 18:38:28.71 & 05:24:35.50 & 13.20 & 0.55 &  --23.1  & 3.1  & 8 & SB2,R \\
29 & 211 & 0957-0382637 & Y & 18:38:39.43 & 05:42:03.92 & 13.25 & 0.13 &  --25.02 & 0.13  & 7 &R,Li\\
30 & 335 & 0954-0380934 & Y & 18:39:39.69 & 05:25:42.38 & 13.28 & 0.18 &  --23.2 & 0.1 & 24 & Li\\
31 & 150 & 0955-0379686 & Y & 18:38:08.03 & 05:35:50.03 & 13.29 & 0.34 &  --23.9  & 1.2  & 22 &SB2,Li\\
32 &  63 & 0955-0378731 & N & 18:37:17.87 & 05:33:09.54 & 13.36 & 0.17 &  --2.6 & 1.5 & 6 &\dots\\
33 &  40 & 0955-0378466 & N & 18:37:06.12 & 05:34:41.92 & 13.47 & 0.15 &  --12.4 & 8 & 16 &\dots \\
34 &  74 & 0954-0378022 & Y & 18:37:24.68 & 05:29:02.98 & 13.61 & 0.29 &  --25.7  & 6.0  & 5 & flares,R,Li\\
35 & 222 & 0952-0376936 & Y & 18:38:41.84 & 05:14:45.42 & 13.63 & 0.29 &  --24.00  & 0.14  & 6 & R,Li\\
36 & 240 & 0952-0377044 & Y & 18:38:47.29 & 05:16:25.21 & 13.68 & 0.38 &  --24.76 & 0.03 & 3 & R,Li\\
37 & 165 & 0954-0378924 & Y & 18:38:13.87 & 05:29:53.84 & 13.70 & 0.26 &  --25.28  & 0.24  &  8& R,Li\\
38 & \dots & 0953-0376947 & Y & 18:37:37.69 & 05:18:37.33 & 13.79 & 0.11 &  --24.55 & 0.03 &  2 & R,Li\\
\noalign{\smallskip}\hline
\end{tabular}

\vspace{1mm}Notes. $^a$no CoRoT photometry. $^b$also observed by CoRoT in LRc05. $^c$R: rotation; Puls: pulsation; SB: spectroscopic binary; EB: eclipsing binary; Li: star with lithium line at $\lambda$670.8\,nm.
\end{flushleft}
\end{table*}

%---------------   F1:  CoRoT light curves
\begin{figure*}[!tbh]
\center
\includegraphics[angle=0,width=90mm]{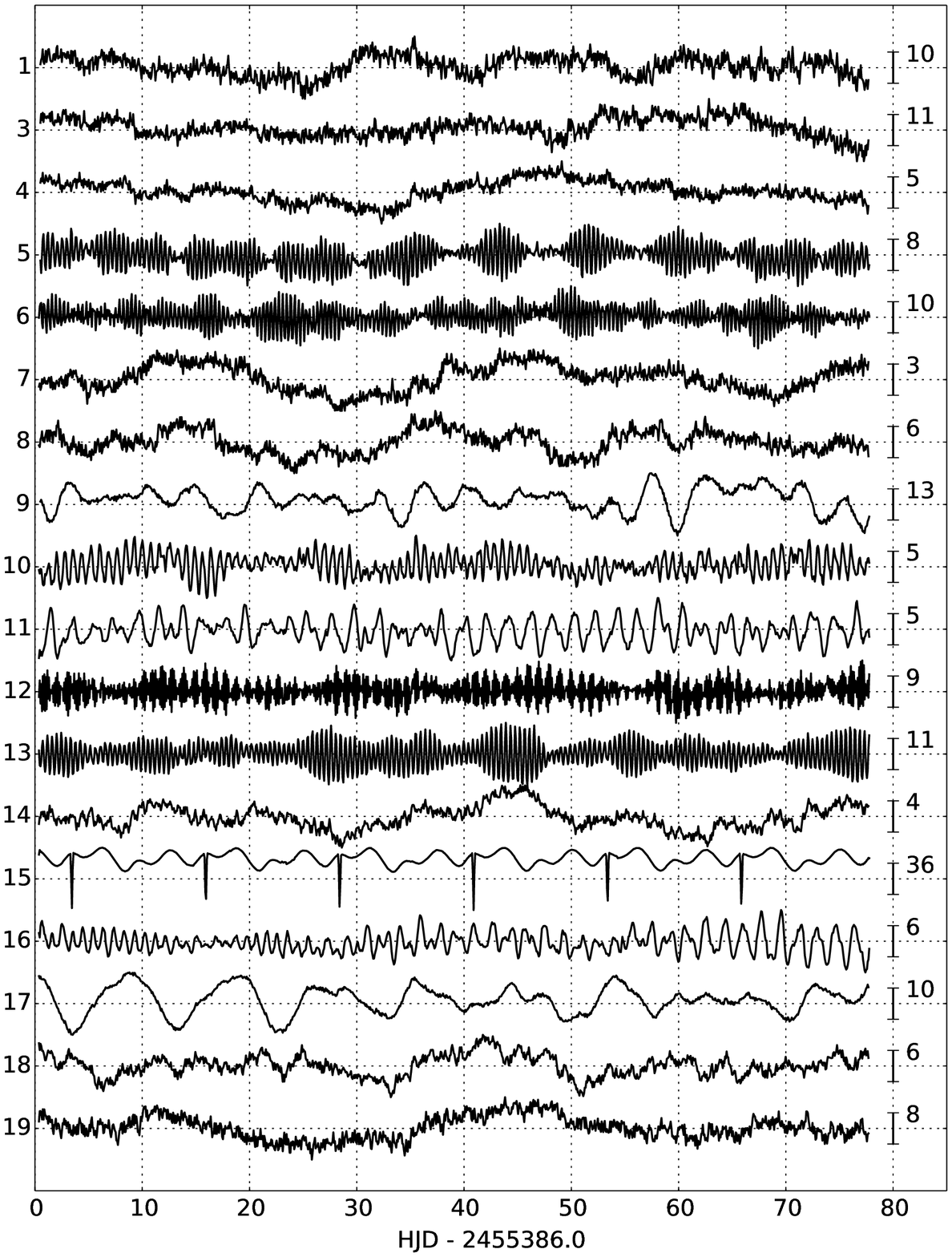}\hspace{1mm}
\includegraphics[angle=0,width=90mm]{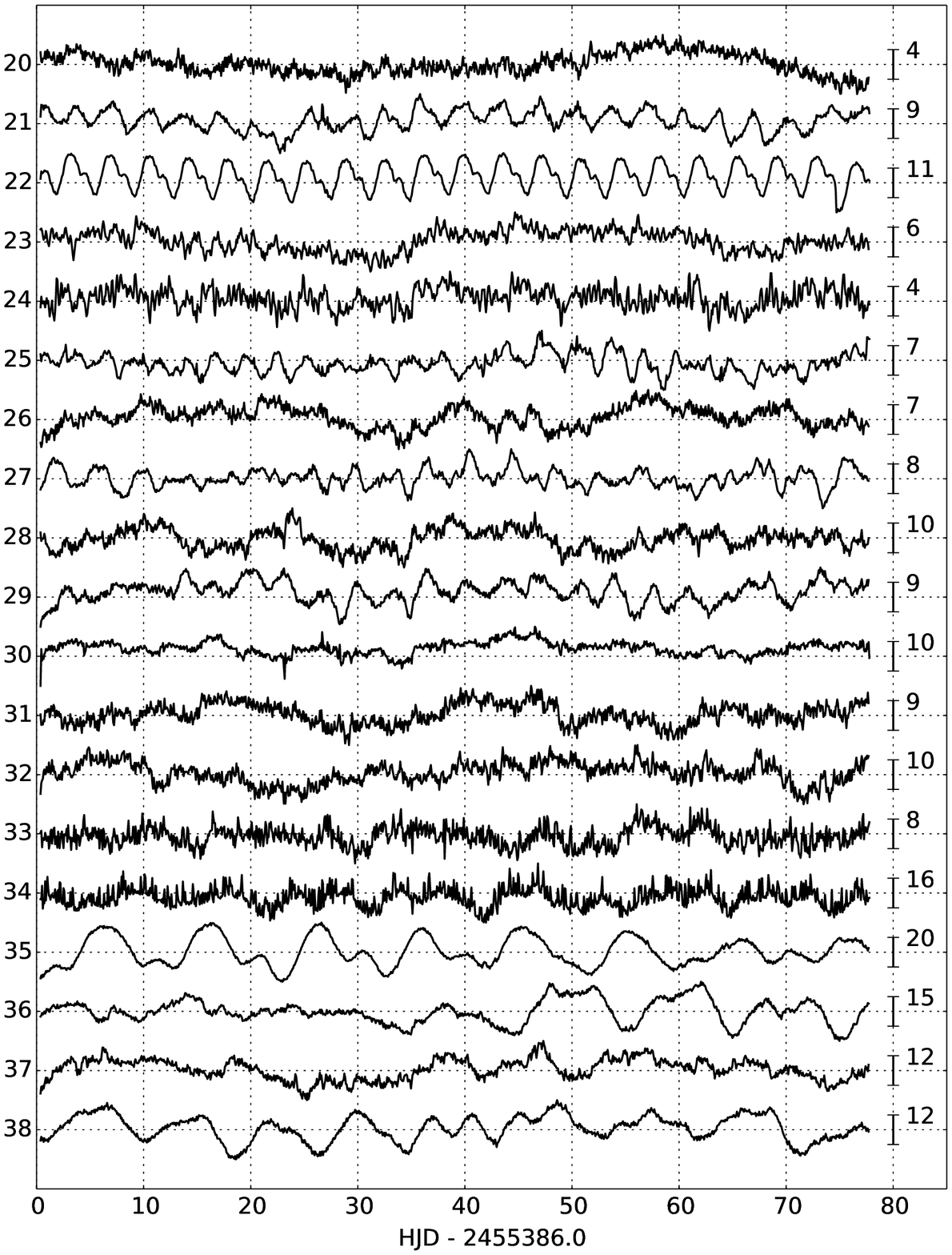}
\caption[ ]{CoRoT light curves for the target stars from Table~\ref{T1}. The running  number on the left side of the plots identifies each star. The numbers on the right side of each plot indicate the scale (varying from star to star) in millimag. Star \#2 is missing because CoRoT photometry could not be obtained for it.
 \label{F1}}
\end{figure*}

\section{Observations and data reductions}\label{S2}

\subsection{CoRoT photometry}

CoRoT observed IC~4756 from 2010 July 8 to September 24 (long run LRc06; 78 days). One of the two CCDs for the exoplanet field was used, and stars between $B\approx$10--14~mag were downloaded. Almost the entire cluster was observable by CoRoT. The cluster's angular diameter has been estimated by Herzog et al. (\cite{herz}) to be around 90\arcmin.

Our full (38-star) CoRoT IC4756 sample is summarized in Table~\ref{T1} and the photometry is shown in Fig.~\ref{F1}. The principal identifier used in this paper is running number in the first column in Table~\ref{T1}. The following columns list the Herzog-Sanders-Seggewiss (HSS) number, the USNO-B1.0 number, a cluster membership flag, followed by the NOMAD coordinates (Zacharias et al. \cite{nomad}).
$V$ magnitudes are taken from NOMAD while the $B-V$ values are from Herzog et al. (\cite{herz}). The next three columns list our radial velocities from STELLA/SES, with an rms value listed when multiple observations were obtained, together with the number of observations. The last column contains miscellaneous notes. The first four target entries in Table~\ref{T1} also have a BD number and/or a number in Kopff (\cite{kopff}), i.e., \#1 = \object{BD+05\,3834} = \object{HSS~73} = Kopff\,43, \#2 = \object{BD+05\,3888} = \object{HSS~314} = Kopff\,139, \#3 = \object{BD+05\,3849} = \object{HSS~138} = Kopff\,68, and \#4 = \object{BD+05\,3844} = \object{HSS~106} = Kopff\,59. Unfortunately, HSS~314 was too bright for CoRoT and was not observed. Targets \#1, 3-5, and 8 were observed in the imagette modus but \#1 was severely overexposed. The flux for target \#11 was contaminated by 41\%\ owing to the presence of another nearby star, and was corrected under the assumption that the close-by star is constant.

The raw LRc06 light curves have 206,470 data points per target, flagged data already having been removed by the CoRoT-N2 pipeline (Deleuil et al.~\cite{del}, Meunier et al.~\cite{meu}). As is common, the background-subtracted white flux was extracted from the light curves, including those from the imagettes. Each light curve was normalized to the median for numerical reasons. In order to increase the S/N ratio and to reduce the number of data points, the light curves were re-sampled from a 32-s cadence to 512\,s. Missing data points were interpolated. This has a negligible effect on further processing, a fact that we verified by checking that it introduces no spurious frequencies. The final set consists of 12,288 equally spaced measurements. The variance within a 512-s bin is used as a preliminary error estimate and amounts to 0.68\,mmag.

The only additional correction performed was the subtraction of a long-term linear trend from the data sets. A polynomial fit is not desirable here because this could cause longer periodic signals to be removed accidentally. The stellar signal is also slightly contaminated by some high-frequency noise, which we removed using a low-pass filter with a cut-off frequency of 28.079\,$\mu$Hz ($\approx$10\,hours).

Three targets were also observed in the prior CoRoT run (LRc05) from 2010 April 8 to July 5, but with only part of the field overlapping with LRc06. One of the three targets (HSS\,348 = \#15 in Table~\ref{T1}) will be analyzed in more detail in a separate paper. However, it is included in this paper's period analysis for the sake of completeness. In LRc06 a major event took place at CoRoT-JD~3785.1243 (HJD 2,455,330.1243) that increased the average flux level by 0.1\%. The three targets that were covered by the two runs were treated individually for each run and concatenated afterward. More details of the data handling and the data reduction process of CoRoT data is given by Weingrill (\cite{wein}).

\subsection{STELLA/SES echelle spectroscopy}

High-resolution spectra were obtained with the \emph{STELLA
Echelle Spectrograph} (SES) at the robotic 1.2-m STELLA-II
telescope in Tenerife, Spain (Strassmeier et al.~\cite{stella},
\cite{malaga}). The SES is a fiber-fed white-pupil echelle
spectrograph with a fixed wavelength format of 388--882\,nm. The
instrument is located in a separated room on a stabilized optical
bench and is fed by a 12-m long 50$\mu$m Ceram-Optec fiber,
corresponding to an entrance aperture of 1.7\arcsec\ on the sky.
This fiber enables a two-pixel resolution of $R$=55,000. The CCD
is an E2V\,42-40 2048$\times$2048 13.5$\mu$m-pixel device.

We obtained integrations with exposure times ranging from 800~sec to
7200~sec, and achieved signal-to-noise (S/N) ratios between 10:1 to
70:1 per resolution element, respectively, depending on target
brightness. Even the lower S/N is still sufficient to obtain a
radial-velocity precision of around 1~\kms\ for slowly rotating
stars (for more details see Strassmeier et al. \cite{orbits}).
Numerous radial velocity standards and stellar comparison targets
were also observed with the same set-up. The best rms radial-velocity
precision obtained over the previous years was 27~\ms\ for high S/N
spectra and stars with narrow spectral lines. The data also included
nightly flat-field exposures, bias readouts, and Th-Ar
comparison-lamp exposures. For details of the echelle data
reduction we refer to Weber et al. (\cite{spie}) and a prior
paper by Strassmeier et al.~(\cite{orbits}). Table~\ref{T1} lists
the number of spectra, $N_{v_{\rm r}}$, the average radial velocity, $v_{\rm
r}$, and the standard deviation of the radial velocity around the
mean, $\sigma_{\rm v}$.

Fifty selected spectral orders are used to determine the stellar
effective temperature, the gravity, the line broadening, and
the metallicity. Our numerical tool PARSES (PARameters from SES; Allende Prieto
\cite{all}, Jovanovic et al. \cite{jov:web}) is implemented as a suite of Fortran
programs in the STELLA data analysis pipeline. All calculations were done using
MARCS model atmospheres (Gustafsson et al. \cite{marcs}) with
the VALD3 line list (Kupka et al. \cite{vald}; and updates on
some specific $\log gf$ values).

%--------------------------   F2:  Color-color diagrams from STELLA/WiFSIP
\begin{figure*}[!tbh]
\includegraphics[angle=0,width=\textwidth,clip]{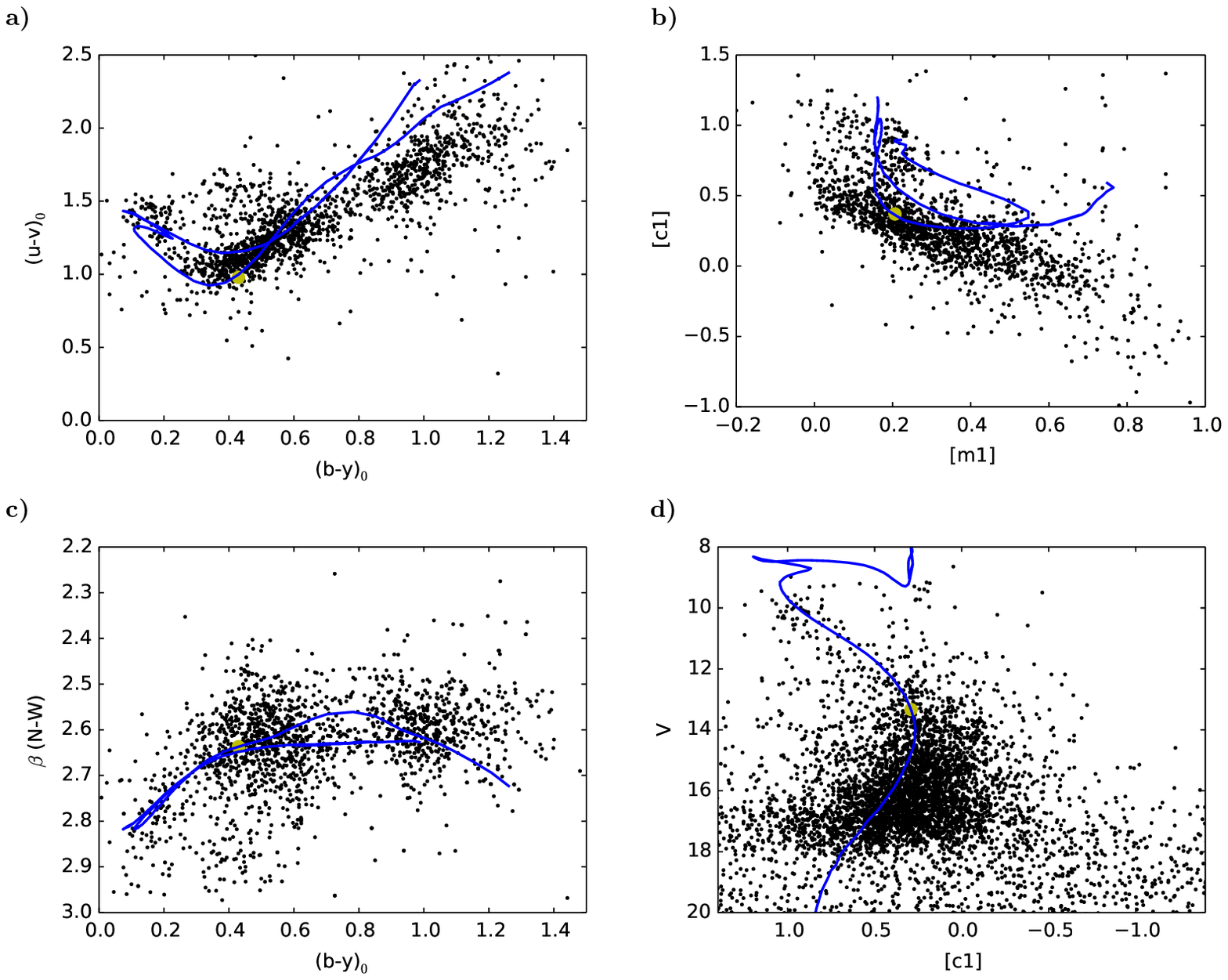}
\caption[ ]{Str\"omgren diagnostic diagrams obtained from deep STELLA/WiFSIP $uvby\beta$ photometry. {\bf a} De-reddened color-color diagram $(u-v)$ versus $(b-y)$; {\bf b} color difference $c_1=(u-v)-(v-b)$ versus the blanketing difference $m_1=(v-b)-(b-y)$; {\bf c} H$\beta$ index versus $b-y$ color; {\bf d} $V$ magnitude versus de-reddened color difference $c_1$. The line is the 890\,Myr isochrone with solar metallicity. The light-gray dot would be a Sun in IC\,4756. We note that in panels a--c only stars with photometric errors smaller than $\approx$0\fm01 are plotted, i.e., typically brighter than 17th magnitude in $y$.
 \label{F2}}
\end{figure*}

\subsection{Str\"omgren and H$\alpha$ STELLA/WiFSIP photometry}

Ground-based CCD photometry was obtained with the \emph{Wide Field
STELLA Imaging Photometer} (WiFSIP) at the robotic 1.2-m STELLA-I
telescope in Tenerife, Spain (Strassmeier et al.~\cite{stella},
\cite{malaga}). Its CCD is a four-amplifier 4096$\times$4096
15$\mu$m-pixel back-illuminated CCD from STA that was thinned and
anti-reflection coated at Steward Imaging Technology Lab. The
field of view with a 3-lens field corrector is an unvignetted
22\arcmin$\times$22\arcmin\ with a scale of 0.32\arcsec/pixel.
The photometer is equipped with 90mm sets of filters with a total of 20 filters in a single filter wheel. For IC\,4756
we employed Str\"omgren-$uvby$, narrow and wide H$\alpha$, and
H$\beta$ filters. Filter curves are available on the STELLA home
page\footnote{www.aip.de/stella}. The WiFSIP narrow-band
filter has 4\,nm FWHM, and the wide-band filter has 16\,nm FWHM.

The observing log is given in Table~\ref{AT1} (in the Appendix),
lists the mean Julian date, the field identification, the
number of CCD frames, the total exposure time on target, the
average FWHM across the field in arcsec, and the average airmass.
Integrations were set to 91s, 29s, 27s, 20s, 160s, 36s, 160s, and 36s for
$u$, $v$, $b$, $y$, H$\alpha$n, H$\alpha$w, H$\beta$n, and H$\beta$w,
respectively. For each filter, five exposures were taken en-bloc, summing to
2795 seconds of exposure per visit.
Six fields (internally called f2, f3, f5, f6, f8, f9) were observed.
Field f2 was only observed in $uvby$. The total number of CCD frames
analyzed was 3,517, but was unevenly distributed among the fields.
The following numbers in parentheses identify the number of visits per field:
f2 (13), f3 (14), f5 (12), f6 (16), f8 (29), and f9 (23).

Coordinates for the field centers are given in the notes to Table~\ref{AT1}
for eq. 2000.0. A mosaic of the entire IC\,4756
field observed by STELLA is shown in Fig.~\ref{Fapp1} in the appendix, where
we also identify the CoRoT targets according to their running numbers in
Table~\ref{T1}. The STELLA observing sequence was based on a merit function that took into account the successful number of previous visits, the current weather conditions, the quality of the previous images, and the duration of the time gaps between previous field visits.

The raw frames were first bias subtracted, flat fielded and illumination
corrected. Astrometry (world-coordinate solution) was performed with the PPMXL
catalog within the WiFSIP data reduction pipeline (Granzer et al. 2015, in
preparation). Then they were bad-pixel-interpolated using a mask obtained from
the flat field, object masked and line fitted to subtract background residuals,
aligned, and average combined per visit (five images per filter), using
standard routines within the IRAF\footnote{IRAF is distributed by the National Optical Astronomy Observatories, which are operated by the Association of Universities for Research in Astronomy, Inc., under cooperative agreement with the National Science Foundation.} environment. For the purposes of combination, the
images were flux scaled according to their exposure time and zero-point
magnitude, and weighted as described in Erben et al. (\cite{erben}) but with an
additional seeing-weight factor inversely proportional to the square of the
full-width-at-maximum, i.e., to the effective area of the (Gaussian)
point-spread function (see F. Masci 2012\footnote{web.ipac.caltech.edu/staff/fmasci/home/mystats/\dots\\ \dots ImCombineWseeing.pdf}).
A threshold was used to reject broad bad pixel features (such as
along the frame borders) and an upper sigma-clipping algorithm was used
to reject cosmic rays and other short transients. The combined images
were re-calibrated astrometrically, and re-sampled onto the same world
coordinate system to account for slight field rotations from night to night and
$\la$4\arcmin\ distortions toward the field corners. Then they were
average combined using the propagated scales and weights to account for the
varying depth, but without sigma-clipping rejection, because image ellipticity and
FWHM vary from night to night.

%--------------------------------   F3:  CMD from STELLA/WiFSIP
\begin{figure*}[!tbh]
\includegraphics[angle=0,width=15cm,clip]{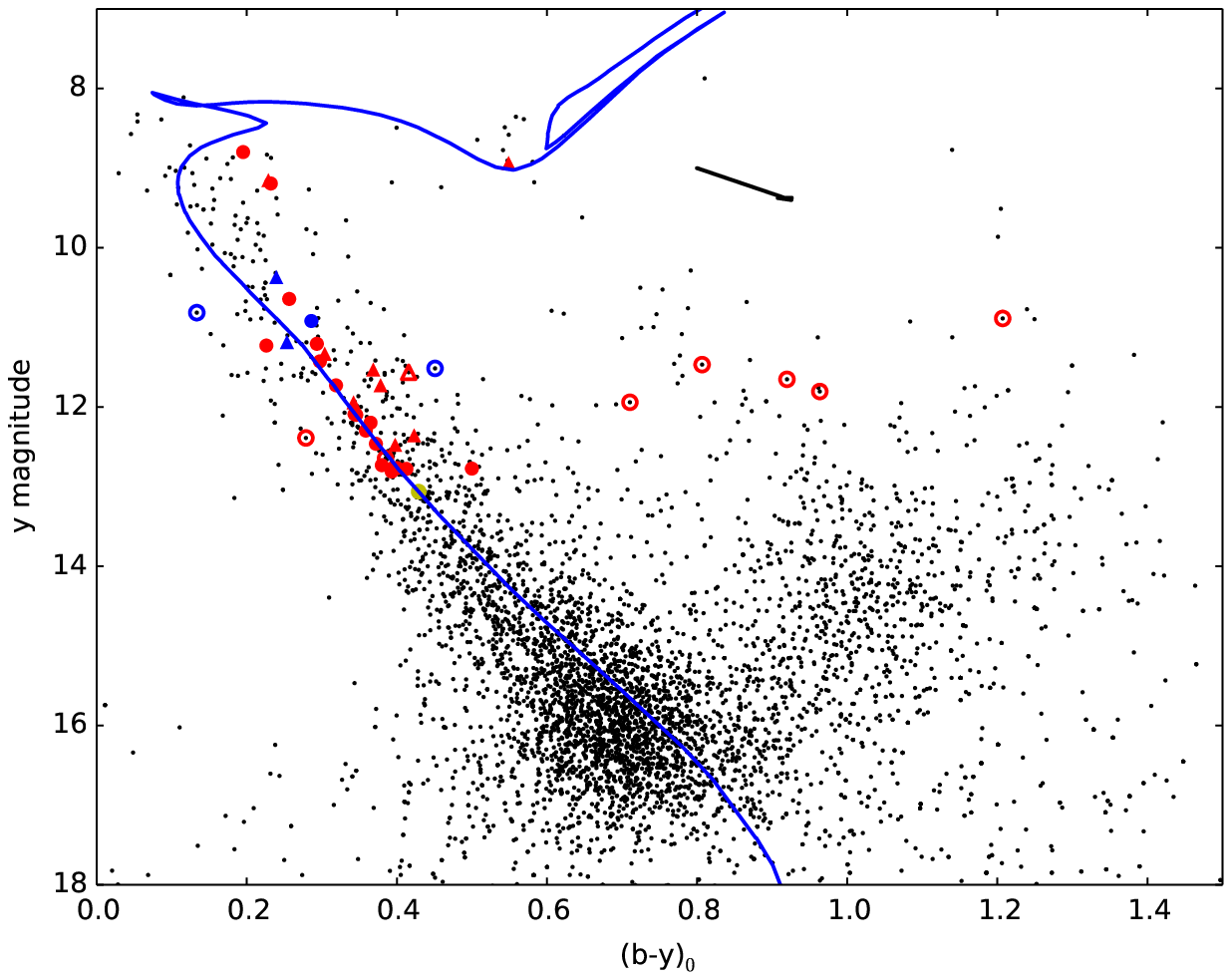}
\caption[ ]{Color-magnitude diagram constructed from STELLA/WiFSIP $uvby$ photometry (small black dots). The best-fit isochrone from PARSEC (Bressan et al. \cite{parsec}) with solar metallicity is overlaid. The filled symbols are the cluster members from our CoRoT sample from Table~\ref{T1}, while the open symbols are the CoRoT non-members. Triangles indicate binaries, blue (dark-gray) symbols moreover indicate the pulsating stars. The single green (light-gray) dot is the Sun. Nine of the CoRoT targets (\#33, 32, 24, 23, 22, 20, 19, 12, 9) appear not to be cluster members. We note that the only cluster giant in our CoRoT sample is HSS\,314 (\#2; red filled triangle on the giant branch), which itself is a spectroscopic binary. The black arrow indicates the reddening vector.
\label{F3}}
\end{figure*}

\section{The IC\,4756 cluster = Melotte 210}\label{S3}

\subsection{Brief literature history}\label{S3.1}

Memberships based on $UBV$ photometry and common proper motions were determined
by Herzog et al. (\cite{herz}) for stars brighter than $B\approx 14$~mag,
following earlier studies by Kopff (\cite{kopff}) and Alcaino (\cite{alc}).
From main-sequence fitting of ten standard clusters, Salaris et al.
(\cite{sal:wei}) revised the age of IC\,4756 to 790$\pm$170~Myr assuming
[Fe/H]=$-0.03$, in good agreement with the 800$\pm$200~Myr obtained later by
Pace et al. (\cite{pace10}).
Radial velocities for 23 members were obtained in the context of their large
program by Mermilliod et al. (\cite{merm08}), who determined an average cluster
velocity of --25.15~\kms \ (see also Mermilliod \& Mayor \cite{merm90} for
earlier results). Herzog et al. (\cite{herz}) presented proper motions and
estimated an approximate cluster angular diameter of 90\arcmin, with 274
probable members.
The mean color excess, from $uvby\beta$ photometry by Schmidt (\cite{sch}),
agreed with the average $BV$-based reddening of $E(B-V)$=0\fm19 from Seggewiss
(\cite{segg}) and Herzog et al. (\cite{herz}), but was shown to vary by up to
 0\fm14 across the area of the cluster. These data were based on just 27 stars.
Twarog et al. (\cite{twa}) also pointed out that the scatter in the
color-magnitude diagram (CMD) makes it difficult to determine its distance.
The distance modulus remained uncertain, but was repeatedly given as
8\fm05$\pm$0.59 (e.g., Schmidt~\cite{sch}), a figure which implies a distance of
$\approx$400~pc. Robichon et al. (\cite{rob}) determined a distance of
330$^{+59}_{-43}$~pc based on Hipparcos data of nine member stars. This figure
needs to be revised to 440$^{+78}_{-57}$\,pc (2.27$\pm$0.34\,mas) based on the
re-reduced Hipparcos data (van Leeuwen \cite{lee}).

Thogersen et al. (\cite{thog}) found a mean [Fe/H] of --0.22$\pm$0.12, based on
eight stars.
Luck (\cite{luck}) determined an average [Fe/H] = --0.03$\pm$0.06 from four
likely members, and also presented abundances of several other elements.
According to Jacobson et al. (\cite{jac}) IC\,4756 shows enhanced values of
[Na/Fe] and [Al/Fe], while the abundances of Fe, O, Si, and Ca are consistent
with their ages and locations in the galactic disk.
They suggest an average cluster [Fe/H] of --0.15$\pm$0.04. More recently,
Pace et al. (\cite{pace10}) found an iron abundance for three late-F giant
stars (HSS 38,42,125) of [Fe/H] = +0.08 and for three mid-G dwarfs (HSS
97,165,240) of [Fe/H] = +0.01, which is basically equal to solar. Ting et al. (\cite{ting}) provided a chemical analysis  for 12 subgiants and measured a cluster metallicity of  $-0.01\pm0.10$. 
Three G dwarfs have lithium abundances between $\log N$(Li) of 2.6 to
2.8~dex with respect to hydrogen ($\log N$=12), and show chromospheric activity
comparable to those of the Hyades and Praesepe stars (Pace et al.
\cite{pace09}). The famous $\delta$~Sct star HD~172189 is not only a
member of the cluster, but also a member of an eclipsing,
double-lined spectroscopic binary system and was studied in great
detail (e.g., Creevey et al. \cite{cree} and references therein).
Their abundance analysis showed [Fe/H] = --0.28 for the $\delta$~Sct
component and +0.4 for the secondary star (as noted by Creevey et
al. their [X/H] values are on average $\approx$0.3\,dex lower than
those from Jacobson et al. \cite{jac}).
Ibanoglu et al. (\cite{iban}) also determined absolute parameters for
HD~172189, determining a distance of 432--453~pc and an age of 890~Myr,
the latter from isochrone fitting to the binary parameters in the H-R
diagram.

An 88-ks ROSAT HRI exposure initially yielded a detection of only a
single cluster member (HSS\,201 = BD+05$^\circ$3863; Randich et al.
\cite{ran}). This target is an A8 star with an X-ray luminosity of
7.3\,10$^{29}$ erg/s, and is not on our cool-star observing list.
Reanalysis of the same exposure by Briggs et al. (\cite{briggs})
yielded four more sources, but with unknown cluster membership.
A study of the rotational velocities in IC\,4756 by Schmidt \&
Forbes (\cite{schm:for}) also focused on A stars as these types
populate the turn-off point. None of these targets matches
the targets studied in the present paper.

\subsection{Cluster membership of the CoRoT stars from radial velocities}

Mermilliod et al. (\cite{merm08}) determined the average cluster velocity from 23 stars to be --25.15~\kms\, with a spread around the mean of approximately $\pm$3~\kms\ peak to valley. This value is on the CORAVEL zero-point system (Famaey et al. \cite{famaey}). Unfortunately, none of their 23 stars is within our CoRoT/SES sample. However, the STELLA/SES zero point was recently determined to be +0.503~\kms\ in the sense STELLA minus CORAVEL (Strassmeier et al. \cite{orbits}). The STELLA/SES cluster mean velocity from 21 members is $-24.0\pm1.3$(rms)\,\kms\ (all velocities in this paper are on the STELLA scale if not otherwise mentioned and given for barycentric Julian date).

Table~\ref{T1} lists our averaged velocities and their standard deviations for the CoRoT stars. With the exception of the few bright targets ($V\approx10^{\rm m}$), all stars are at the S/N-ratio limit of the 1.2-m STELLA SES system. For the stars in the $V$ range 10$^{\rm m}$--12$^{\rm m}$, we obtain radial-velocity residuals of typically between 50--300~\ms\, while for stars fainter than 12$^{\rm m}$ this increases to $\approx$0.5--1~\kms, always depending on the line broadening. Therefore, we emphasize that our systematic synthetic-spectrum fitting of all SES spectra with PARSES also comes with larger than typical errors because of the low S/N. Fifty echelle orders of each SES spectrum are selected for the fitting with the following four free parameters; $T_{\rm eff}$, $\log g$, $[$Fe/H$]$, and rotational line broadening $v\sin i$ (for a more detailed description see Strassmeier et al. \cite{orbits} or Jovanovic et al. \cite{jov:web}).  Additional spectral information, if reliable due to the low S/N ratio, is given in the individual notes.

Five CoRoT targets (\#20, \#19=HSS\,185, \#23=HSS\,89, \#24=HSS\,198, \#9=HSS\,356) in the color-magnitude diagram in Fig.~\ref{F3} do not fit into the cluster sequence at all. These stars appear too red in $b-y$ by 0\fm3--1\fm0 for their brightness and are consequently judged to be either non-members or to be severely affected by intervening inhomogeneous dust clouds. Because their radial velocities are also significantly off the cluster mean, we are confident that none of them is a cluster member. Four more target are less discrepant ($\pm\approx0\fm1$) to the CMD main sequence but still show far too low a radial velocity to be considered cluster members. These nine stars are indicated in Fig.~\ref{F3} as open symbols.
See the individual notes in the appendix.

\subsection{Reddening, distance, metallicity}

The dark nebula LDN\,630 is very close to IC\,4756 at RA\,18~38.9 and DEC+06~03 (Dutra \& Bica \cite{LDN}). Although not exactly in the line of sight, some of the peripheral regions of the nebula affect the colors of IC\,4756 stars in the WiFSIP frames. The dust maps of Schlegel et al.~(\cite{sch:fin}) indicate a very large \emph{average} reddening of $E(B-V)\approx 0\fm75$ ($A_V$=2\fm34) at the central position of IC\,4756 compared to the $E(B-V)$=0\fm19 obtained by Herzog et al.~(\cite{herz}) and $0\fm22$ by Schmidt (\cite{sch}). The NASA/IPAC dust maps show a patchy absorption pattern across the cluster area which is also nicely seen in the WISE images\footnote{See the composite with an optical image by J.~Thommes at http://www.jthommes.com/Astro/IC4756.htm} at 4.6$\mu$m and 12$\mu$m. Our $uvby$ photometry is compatible with the $0\fm19$ value. With a reddening of $0\fm75$ no agreement between our photometric data and standard evolutionary tracks can be achieved. This suggests that most of the dust in the IPAC and WISE data is located behind IC\,4756.

The only prior Str\"omgren photometry is from Schmidt (\cite{sch}). He noted that the calibration of his four-color photometry in terms of an absolute zero point was not always applicable to IC\,4756, most likely because of foreground dust lanes likely related to the outskirts of LDN\,630. We adopted standard stars from the compilation of Hauck \& Mermilliod (\cite{hauck}) and found only a total of eight stars within our  field of view suitable for a zero-point calibration of the STELLA/WiFSIP data. The standard $uvby\beta$ transformation was applied by following the procedure given in Balog et al. (\cite{balog}, and references therein). The equations are given in Appendix~A. The best value for the reddening is $E(b-y)$=0\fm16$\pm$0.02 and the distance modulus $V-M_{V}$=8\fm02$\pm$0.08. This equates to $E(B-V)$=0\fm22 in the Johnson system. For the calibration of our measured $\beta$-indices the eight stars were only of marginal use because of their strong correlation with $(b-y)$. The final results for the CoRoT targets are listed in Table~\ref{AT2} in the appendix.

Fig.~\ref{F2}a shows the de-reddened $u-v$ vs. $b-y$ color-color relation. Only stars with photometric errors below $0\fm01$ are plotted, i.e., stars brighter than $V\approx 17^{\rm m}$. Some of the scatter is also due to the lack of very cool standard stars in our color-color calibration. For M dwarfs ($0\fm8<(b-y)_0<1\fm4$), we see a large difference with respect to the solar-metallicity isochrone in the sense that the M dwarfs appear too bright in the ultraviolet by 0\fm2. However, almost all of these stars are non-member foreground stars.

Fig.~\ref{F2}b plots the de-reddened color difference $c_1$ as a measure of the discontinuity of the continuous hydrogen absorption vs. the de-reddened blanketing difference $m_1$ for the same sample of stars as in Fig.~\ref{F2}a. For our cool stars, $c_1$ is actually a measurement for absolute magnitude and surface gravity while $m_1$ is a measure for metallicity. The homogeneous pattern in the figure suggests that these stars are indeed mostly cluster members while the stars significantly above and below the pattern are likely foreground or background stars. The relations in Holmberg et al. (\cite{holm}) are used to determine individual metallicities for the validity range $0\fm2<(b-y)_0<0\fm6$. The average metallicity from 19 members of $-0.19\pm0.3$  (Table~\ref{T2}) is basically consistent with solar metallicity. However, its scatter of 0.3\,dex indicates the inhomogeneity of individual stars. The average metallicity of 9 members from spectrum synthesis is $-0.13\pm 0.09$, while the isochrones give equally good fits for a large range of metallicities peaking near solar values. While the grand non-weighted average from spectroscopy and photometry is $-0.08\pm0.06$ (rms), the average from the CMD-fitted metallicities is $-0.03\pm 0.02$ (rms). We consider both values consistent with solar metallicity.

Fig.~\ref{F2}c employs the H$\beta$ index defined as the brightness difference in the narrow-minus-wide filters for the same sample as in Fig.~\ref{F2}a. It basically measures temperature for the stars cooler than spectral type $\approx$A2 ($b-y\geq 0\fm03$). While $b-y$ also defines the temperature scale but is not free of blanketing, the \Hbeta\ index is additionally independent of interstellar reddening. We again apply the relations in Holmberg et al. (\cite{holm}) to determine effective temperatures and absolute magnitudes.

Fig.~\ref{F2}d shows $V$ brightness vs. the de-reddened color difference $c_1$. It is an observational H-R diagram. We note that we transformed $y$ to $V$ with the relation in Olsen (\cite{olsen94}) for more convenient comparison with other clusters, and also extended the plotting range to $V\approx 20^{\rm m}$. Almost all of the stars in the range 18--20$^{\rm m}$ are background stars and only one of the fields has been plotted (because the region would appear black otherwise). We note that a Sun in IC\,4756 would appear at $V\approx 14^{\rm m}$ (gray dot). The lack of bright evolved stars is compatible with the intermediate age of 890\,Myr. The isochrone very well fits the 10 giants in Fig.~\ref{F3} that were already identified by Schmidt (\cite{sch}) and which are also part of the WiFSIP data.

%------------------------------ Table 2:  Metallicity and age results
\begin{table}[!tbh]
\begin{flushleft}
\caption{Metallicity and age results.}\label{T2}
\begin{tabular}{llll}
\hline \noalign{\smallskip}
Method & [Fe/H] & Age & Notes \\
\noalign{\smallskip}\hline \noalign{\smallskip}
Spectra of member stars & --0.13$\pm$0.09 & \dots & 9 stars \\
$c_1$ vs. $m_1$         & --0.19$\pm$0.3 & \dots & 19 stars\\
CMD $y$-$b-y$           & --0.05$\pm$0.20 & 890$\pm$70 & isochrone\\
CMD $y$-$u-b$           & --0.03$\pm$0.21 & 620$\pm$340 & isochrone\\
CMD $y$-$v-y$           & --0.01$\pm$0.22 & \dots & isochrone\\
\noalign{\smallskip}\hline \noalign{\smallskip}
Non-weighted average & --0.08$\pm$0.06 & & \\
CMD-weighted average & --0.03$\pm$0.02 & & \\
\noalign{\smallskip}\hline
\end{tabular}
\end{flushleft}
\end{table}

\subsection{Cluster age}

Our main CMD from the STELLA WiFSIP photometry is shown in Fig.~\ref{F3}. The error level in $y$ is on the order of mmag for stars brighter than 13th magnitude and up to $0\fm01$ for stars near 16th magnitude. The error level is larger by up to a factor of two for $b$ and $v$ as well as for $u$. The small dots in Fig.~\ref{F3} indicate the WiFSIP stars and the big dots indicate the CoRoT stars. For the plot, we did not remove any of the non-members from the CoRoT sample but number specific outliers for identification purposes.

A main-sequence turn-off age is determined by fitting an isochrone from PARSEC (Padova models; Bressan et al. \cite{parsec}). PARSEC is a stellar evolution code with up-to-date physics for post- and pre-main-sequence stars, with particular emphasis on chemical mixture and its relation with physical properties like the opacities and the equation of state. Its abundance scale has been adapted to the new solar chemical abundances with a solar metallicity of $Z$=0.0152. Isochrones are provided for several photometric systems and for a large range of ages and metallicities. Colors are based on calibrations from ATLAS-9 spectra (Castelli \& Kurucz \cite{atlas-9}). Small but systematic differences are not surprising and are partly due to the isochrone calibration itself (see Clem et al. \cite{clem}).

No targets fainter than $y$=16\fm0 were included in the isochrone fit owing to the increasing contamination with background stars. Fig.~\ref{F3} shows the best-fit isochrone with an overshoot parameter of 0.5, a formal cluster metallicity of [Fe/H]= $-0.05$ (which we deem still solar), a distance modulus $V-M_V$=8\fm02, and for a reddening of $E(b-y)$=0\fm16. No particular isochrone gives a perfect fit in all colors. This is not unexpected given the small number of members and the lack of a well-developed giant branch at that age. The isochrone fit to the CMD from $by$ data show the least rms error, and is consistent with the CMD from $ub$. The $vy$-CMD fit does not converge and thus would blur the age determination and is rejected. We note that the external errors of our CCD photometry for stars fainter than 16th magnitude would add to the general uncertainty and these stars are also not considered here. Such challenges (and other observational ones) in determining ages of open clusters were recently reviewed by Jeffery (\cite{jeff}) and we refer to this paper for further discussion. We infer a best-constrained age of 890$\pm$70~Myr from our $by$ data. The turn-off mass for IC\,4756 is near 1.8\,M$_\odot$ ($b-y$$\approx$0\fm1; mid-A stars). Table~\ref{T2} summarizes the various results from the color-magnitude diagrams.

% ---------------------- Table 3
\begin{table}[tbh]
\begin{flushleft}
\caption{Photometric periods from the CoRoT data set.}\label{T3}
\begin{tabular}{lllll}
\hline \noalign{\smallskip}
No. & HSS & CoRoT & Period$^a$ & Amplitude$^b$\\
    &     & I.D.  & (d)    & (mag)\\
\noalign{\smallskip}\hline \noalign{\smallskip}
 1 & 73 &104884361 & \dots  & \dots \\ %  HSS73: 1 high peak in ACF: noise? shouldn't we remove ths one?
 3 &138 &105005438 & \dots  & \dots \\ %  HSS138: no signal
 4 &106 &104937204 & \dots  & \dots \\ %  HSS106: no signal
 5 &108 &104940248 & 0.43834$\pm$0.000026 & 0.046$\pm$0.018\\ %  HSS108: pulsator
 6 &113 &104957100 & 0.35311$\pm$0.000011 & 0.049$\pm$0.030\\ %  HSS113: pulsator
 7 &112 &104956719 & \dots  & \dots \\ %  HSS112: FFT: 14d period, unclear in ACF (40d?)
 8 &107 &104940055 & 10.5$\pm$0.5  & $\approx$0.01 \\ %  HSS107: 11 days in ACF, 10 days in FFT, amp=0.01
 9 &356 &105337115 & 4.25443$\pm$0.00082& 0.050$\pm$0.038\\ %  HSS356: highly variable LC, diff rot in ACF
10 &241 &105167156 & 0.79648$\pm$0.00013 & 0.018$\pm$0.011\\ %  HSS241: pulsator
11 &269 &105209083 & 2.03383$\pm$0.00075 & 0.031$\pm$0.010\\ %  HSS269: clear signal
12 &208 &105137398 & 0.154691 & 0.061$\pm$0.018\\ %  HSS208: pulsator, period error?!
13 &270 &105210380 & 0.440273$\pm$0.000027 & 0.083$\pm$0.032\\ %  HSS270: pulsator
14 & 57 &104857708 & 10.792$\pm$0.018 & 0.009$\pm$0.005\\ %  HSS57: 2 periods (10.7 and 1 day?)
15 &348 &105328054 & 6.2537$\pm$0.0037 & $\approx$0.2\\ %  HSS348: eclipsing binary at 2P
16 & 95 &104925676 & 1.6828$\pm$0.0011 & 0.032$\pm$0.020\\ %  HSS95: clear differential rotation
17 &174 &105070416 & 9.2582$\pm$0.0066 & 0.071$\pm$0.022\\ %  HSS174: clear signal, diff. rot. in ACF
18 &183 &105087900 & 5.4350$\pm$0.0012: & 0.005$\pm$0.005\\ %  HSS183: messy signal, no ACF
19 &185 &105095195 & \dots  & \dots \\ %  HSS185: no signal
20 &\dots &104963014 & \dots  & \dots \\ %  no HSS: no signal
21 &194 &105112517 & 3.5287$\pm$0.0026 & 0.017$\pm$0.013\\ %  HSS194: clear ACF, diff. rot.
22 &158 &105037080 & 3.6655$\pm$0.0009 & 0.113$\pm$0.017\\ %  HSS158: binary
23 & 89 &104917505 & \dots  & \dots \\ %  HSS89: no signal detected
24 &198 &105121990 & 7.1155$\pm$0.0072 & 0.006$\pm$0.008\\ %  HSS198: diff.rot. in ACF
25 &209 &105136898 & 3.126$\pm$0.005 & 0.018$\pm$0.017\\ %  HSS209: messy signal, no ACF, <1d pulsation?
26 &259 &105191352 & 5.925$\pm$0.01 & 0.005$\pm$0.005\\ %  HSS259: messy LC, no clear ACF or FFT
27 &284 &105227823 & 3.87459$\pm$0.00066 & 0.031$\pm$0.013\\ %  HSS284: low diff. rot.
28 &189 &105100604 & 7.2:  & 0.006$\pm$0.007\\ %  HSS189: messy signal,no ACF, shouldn't we remove ths one?
29 &211 &105137060 & 3.4198$\pm$0.0017 & 0.018$\pm$0.013\\ %  HSS211: spurios peak at 6d in ACF, prob. diff. rot.
30 &335 &105314448 & \dots  & \dots \\ %  HSS335: no signal
31 &150 &105030838 & \dots  & \dots \\ %  HSS150: ACF, LC: 25d, FFT ~12d
32 & 63 &104866124 & \dots  & \dots \\ %  HSS63: no signal detected
33 & 40 &104831451 & \dots  & \dots \\ %  HSS40: no signal detected
34 & 74 &104885763 & 10.145$\pm$0.008 & 0.025$\pm$0.028\\ %  HSS74: flaring star
35 &222 &105145327 & 9.847$\pm$0.008 & 0.131$\pm$0.036\\ %  HSS222: clear signal, 2 evolving spot groups at opposite longitudes
36 &240 &105165044 & 11.420$\pm$0.012 & 0.047$\pm$0.023\\ %  HSS240: diff. rot.
37 &165 &105050653 & 8.589$\pm$0.011 & 0.014$\pm$0.010\\ %  HSS165: messy LC, amp underestimated
38 &\dots &104928150 & 8.7755$\pm$0.0067 & 0.057$\pm$0.022\\ %  no HSS: clear signal, diff. rot. in ACF
\noalign{\smallskip}\hline
\end{tabular}
\end{flushleft}
Note: $^a$Periods of less than one day are due to stellar pulsation, all others except \#9 (see individual notes) are due to rotation. $^b$Time-averaged amplitudes.

\end{table}

\section{Global rotation periods from CoRoT data}\label{S4}

\subsection{Periodogram technique and application}

The detection of stellar rotation periods was accomplished in three different stages. The first stage was the calculation of the Fourier autocorrelation function (ACF). As demonstrated by Aigrain et al.~(\cite{aig}) and McQuillan et al. (\cite{mam}) the first side lobe of the autocorrelation-spectrum serves as a rough estimate for the stellar rotation period and an initial error estimate. While this method proves to be very reliable for periods of several days, it is often difficult or impossible to determine a local maximum below one day lag. The second step consists of calculating the Lomb-Scargle (L-S) periodogram. Lomb (\cite{lomb}) proposed to use least-squares fits to sinusoidal curves and Scargle (\cite{sca}) extended this by deriving the null distribution for it. We use the formulation implemented by Press et al. (\cite{press}) and compute false alarm probabilities following Frescura et al. (\cite{fres}). The previously determined period from the autocorrelation is taken as a constraint. In case of a mismatch of both periods, a verification is done visually. As a last step a sinusoid is fitted to the light curve with the previously determined period in order to obtain a time-average amplitude. The final error estimate for the period was obtained through Bayesian testing; the signal with the best-determined period was removed from the light curve and an artificial signal at random periods was injected and its period then measured with the same procedure.

\subsection{Rotation vs. pulsation and binarity}

The resulting periods and amplitudes are given in Table~\ref{T3}. The derived periods range from 0.15\,d for star \#12 to 11\,d for star \#36. The average amplitudes in Table~\ref{T3} are only guidelines because amplitude variations are the rule and typically span a factor of 2--3 over the time-span of the observations. We note again that the light curves were normalized and linear trends removed. This effectively prevents any statement regarding multi-periodic behavior with periods longer than $\approx$1/2 of the length of the data set ($\approx$40~days for LRc06).

Five systems (\#5, 6, 10, 12, 13) are non-radial pulsators with rather short fundamental periods ($P < 0.8\,d$). These are obvious in the light curves shown in Fig.~\ref{F1}; these systems exhibit many more periods than just the one listed in Table~\ref{T3}. Two additional systems (\#15, 22) show very stable light variations over the time of observation and are classified as close binaries. Star \#15 (HSS\,348) additionally shows narrow eclipses and is therefore actually a triple system.

% --------------------------------- Table 4
\begin{table}[tbh]
\begin{flushleft}
\caption{CoRoT targets with signatures of differential rotation (see also Fig.~\ref{F-diff}).}\label{T4}
\begin{tabular}{llllll}
\hline \noalign{\smallskip}
No. & HSS & $P_\mathrm{rot}$ & $\Delta\Omega/\Omega$ \\
    &     & (d)              & (--) \\
\noalign{\smallskip}\hline \noalign{\smallskip}
16 & 95 & 1.681 $\pm$ 0.0003 & 0.04362 $\pm$ 0.00001 \\
17 & 174& 9.248 $\pm$ 0.005 & 0.07904 $\pm$ 0.00009 \\
21 & 194& 3.332 $\pm$ 0.001 & 0.05384 $\pm$ 0.00003 \\
25 & 209& 3.134 $\pm$ 0.002 & 0.1493 $\pm$ 0.0001 \\
27 & 284& 3.874 $\pm$ 0.002 & 0.10845 $\pm$ 0.00009 \\
34 & 74 &10.12 $\pm$ 0.025 & 0.1400 $\pm$ 0.0005 \\
\noalign{\smallskip}\hline
\end{tabular}
\end{flushleft}
\end{table}

\subsection{Evidence of differential rotation}

The analysis of differential rotation in the light curves is based on two tools, basically the autocorrelation function already mentioned and a Fourier analysis with a Fast Fourier Transform (FFT). Narrow frequencies or periods cannot be resolved at lower frequencies in the power spectrum, but are visible as a beat frequency in the ACF. Usually, two narrow periods are identifiable after five to ten cycles in the ACF as two separate peaks, where their maxima can be determined. We identify them as the side-lobes in the FFT and use a least-squares spectral decomposition (LSSD) to find the best period for the targets with side lobes. These periods are listed in Table~\ref{T4} and supersede the periods in Table~\ref{T3} from the ACF despite that the error bars are in some cases larger.

\begin{figure}[!tbh]
\includegraphics[angle=0,width=86mm]{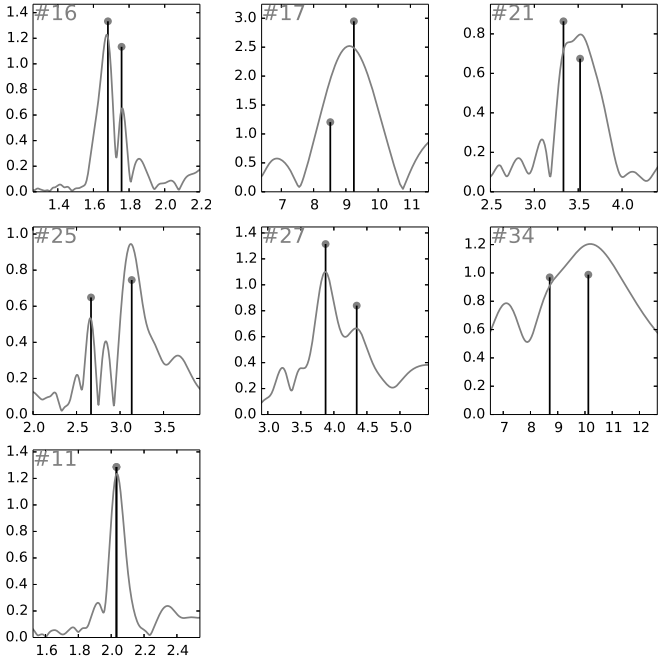}
\caption[ ]{Stars with evidence of differential surface rotation. Shown are relative amplitude vs. period in days of the stars in Table~\ref{T4}. The main periods are indicated with dots and dashes. Star \#11 is a comparison star where no differential rotation is evident.
 \label{F-diff}}
\end{figure}

Six targets show a frequency splitting in the L-S periodogram at the rotational frequency. Fig.~\ref{F-diff} shows the frequency splitting for the six stars and compares them with a representative target (\#11) where no splitting is detected. We plot period on the x-axis in Fig.~\ref{F-diff} in order to compare with other periods in Table~\ref{T4}. We interpret the splitting to be due to differential surface rotation and convert it into radians per day. Numerical values in terms of $\Delta\Omega/\Omega$ are between 0.044 and 0.15 and compare to the solar value of 0.2. Its errors are the quadratically summed relative errors from the individual periods. Three additional targets (\#9, 36, 38) had unusually extended side lobes so that we tend to interpret them to be dominated by shot-term spot evolution rather than differential rotation. It is worth mentioning that the six targets in Table~\ref{T4} show always more than just one side lobe which we interpret as evidence of spots at more than just two latitudes.

%------------------------------   F4:  Diff Rot Periodograms

%------------------------------   F5: Radial velocity curves
\begin{figure}
\includegraphics[angle=0,width=86mm]{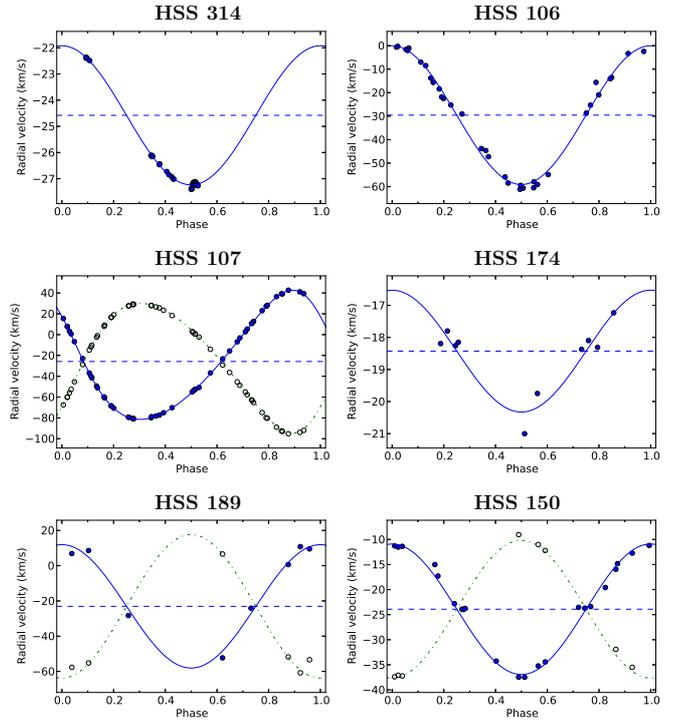}
\caption[ ]{Radial velocity variations and orbital solutions for the six stars in Table~\ref{T5}. We note the preliminary character for the orbits for HSS\,314, 174, and 189.
 \label{F5}}
\end{figure}

\section{Orbital elements}

Eleven of our bona fide IC\,4756 members turned out to be spectroscopic binaries (identified in Table~\ref{T1}). One binary has been known previously (HSS\,314 = \#2); the others are new detections. For six targets, we have enough velocity measurements to principally solve for the usual orbital elements of a spectroscopic binary (three of them being preliminary owing to the poor phase coverage). Radial velocities (RVs) were determined from a simultaneous cross-correlation of 62 \'echelle orders with a synthetic spectrum matching the target properties as far as known (see Weber \& Strassmeier \cite{capella} for a detailed description). We apply the general least-squares fitting algorithm {\em MPFIT} (Markwardt \cite{mpfit}) and refer to Strassmeier et al. (\cite{orbits}) for computational details. Table~\ref{T5} summarizes the orbital elements for these six  targets and Fig.~\ref{F5} compares the observations with the predicted elements.

% ------------------------- Table 5: SBs with orbits
\begin{table*}
\caption{STELLA orbital solutions. Column ``rms'' denotes the quality of the orbital fit for a measure of unit weight in \kms .}\label{T5}
\begin{tabular}{llllllllllll}
\hline \noalign{\smallskip}
\# & HSS & $P_{\rm orb}$ & $T_{\rm Per}^a$ & K  & $\gamma$ & $e$ & $\omega$ & $a_1\sin i$ & $f(M)$ & rms & N \\
   & & (days) & (HJD 245+)      & (\kms) & (\kms)   & & (deg) & (10$^6$ km) & & (\kms) & \\
\noalign{\smallskip}\hline \noalign{\smallskip}
2 & 314 & 2487 & 10507 & 2.6 & --24.6 & 0.0 & \dots & 91  & 0.0048  & 0.06 & 39\\
  &     & $\pm$170 & $\pm$270 & $\pm$0.1 & $\pm$0.1 &  &  & $\pm$7 & $\pm$0.0007  &  & \\
4 & 106 & 23.082 & 5945.2 & 29.5 & --29.6 & 0.0 & \dots & 9.37  & 0.062  & 2.1 & 34\\
     &  & $\pm$0.007 & $\pm$0.2 & $\pm$0.5 & $\pm$0.4 &  &  & $\pm$0.17 & $\pm$0.003& & \\
8$^b$ & 107 & 11.6504 & 5971.603 & 62.03 & --25.81 & 0.174 & 53.6 & 9.785  & 1.110 & 0.44 & 52 \\
     &  & $\pm$0.0003 & $\pm$0.018 & $\pm$0.08 & $\pm$0.04 & $\pm$0.001 & $\pm$0.4 & $\pm$0.013 &$\pm$0.005& & \\
     &  &  & & 62.3 &  &  & & 9.827  & 1.105 & 0.38 & 52 \\
     &  &  & & $\pm$0.13 &  &  & & $\pm$0.020 &$\pm$0.004& & \\
17$^c$ & 174 & 37.31 & 5789.7 & 1.9 & --18.4 & 0.0 & \dots & 0.97  & 0.0  & 0.37 & 10\\
  &     & $\pm$0.52 & $\pm$8.7 & $\pm$0.3 & $\pm$0.15 & & & $\pm$0.15 &  &  & \\
28$^b$ & 189 & 110.0 & 6446.1 & 35.0 & --23.1 & 0.0 & \dots & 53.0  & 2.66 & 3.1 & 8 \\
     &  & $\pm$1.5 & $\pm$0.9 & $\pm$2.1 & $\pm$1.3 &  &  & $\pm$3.3 & $\pm$0.26  & & \\
     &  & & & 40.7 &  &  &  & 61.6  & 2.3 & 3.6 & 6 \\
     &  & & & $\pm$1.6 & & & & $\pm$2.5 & $\pm$0.3  & & \\
31$^b$ & 150 & 19.205 & 6402.34 & 13.01 & --23.91 & 0.0 & \dots & 3.43  & 0.0196  & 1.2 & 22 \\
     &  & $\pm$0.008 & $\pm$0.13 & $\pm$0.1 & $\pm$0.09 &  &  & $\pm$0.03 &         $\pm$0.0005 &  &  \\
            &       &      &       & 13.74 &  & & & 3.63  & 0.0185  & 0.4 & 8 \\
     &  &  &  & $\pm$0.15 & & & & $\pm$0.04 &  $\pm$0.0004 &  & \\
\noalign{\smallskip}\hline
\end{tabular}

\vspace{1mm} $^a$Time of periastron, or ascending node for circular orbits.\\
$^b$SB2. Column $f(M)$ lists then $M_1\sin^3 i$ and $M_2\sin^3 i$ in solar masses.\\
$^c$SB2 but only SB1-orbit possible owing to unfortunate phase coverage.
\end{table*}

Six of the 11 targets are double-lined spectroscopic binaries. One target, HSS\,106 (\#4), appears double lined at first glance but one line system appears constant while the other varies sinusoidally (see individual notes). It is thus a single-lined spectroscopic binary. HSS\,174 (\#17) is a similar case with a broad-lined spectrum whose RVs are used to determine a preliminary SB1 orbit. Its narrow-lined spectrum appears to be constant. Another target, HSS\,108 (\#5), shows RV variations of up to $\approx$18\,\kms\ but is also a pulsating star and the RVs are only marginally consistent with orbital variations as its pure cause. We computed a hypothetical orbit from the current data but list it only in the individual notes for guidance.

\section{\Halpha\ photometry and spectroscopy}

\subsection{Photometric filter definition}

In order to access a magnetic activity tracer in as many cluster stars as possible, we performed \Halpha\ wide- and narrow-band photometry with STELLA, where WiFSIP is equipped with Narrow-band (FWHM 4\,nm) and wide-band (FWHM 16\,nm) \Halpha\ filters.  A total of 356 CCD frames in all fields were obtained and stacked to create a deep \Halpha\ image across the cluster reaching an equivalent $V$ brightness of $\sim$20th mag. We follow the annotations of Herbst \& Layden (\cite{herbst}) and determine a $N-W$ (narrow minus wide) $\alpha$ index from this photometry from
\begin{equation}\label{eq1}
\alpha (N-W) = -2.5 \log N/W .
\end{equation}

In Fig.~\ref{F6}a, we plot the internal errors for approximately 7,000 targets from the stacked CCD image. Their formal standard errors range from 0.1--0.2~mmag for the brightest targets ($V$ brighter than 10\,mag) to about 0\fm1 for the faint limit near $V$$\approx$20\,mag.

\subsection{An activity H-R diagram for IC\,4756}

At the age of $\approx$900\,Myr, we do not expect much magnetic activity left from the early stage at the ZAMS. We know from the Hyades that the magnetic brake has already efficiently slowed down the rapid rotators at its 625\,yrs age and thereby effectively switched off magnetic activity. The 11 targets in IC\,4756 with large amounts of surface lithium are all verified to be definite cluster members. Surface lithium is a sign of either youth and/or some rotation-induced extra mixing. Unsurprisingly, 3 of the 11 Li stars are close binaries while two additional ones are possible binaries with high rotation rates. None of the Li stars is a giant.

%------------------------------   F6:  err vs. V or y  und  alpha vs. b-y
\begin{figure}[!tbh]
{\large \bf a) \Halpha\ standard errors}\\

\includegraphics[angle=0,width=86mm]{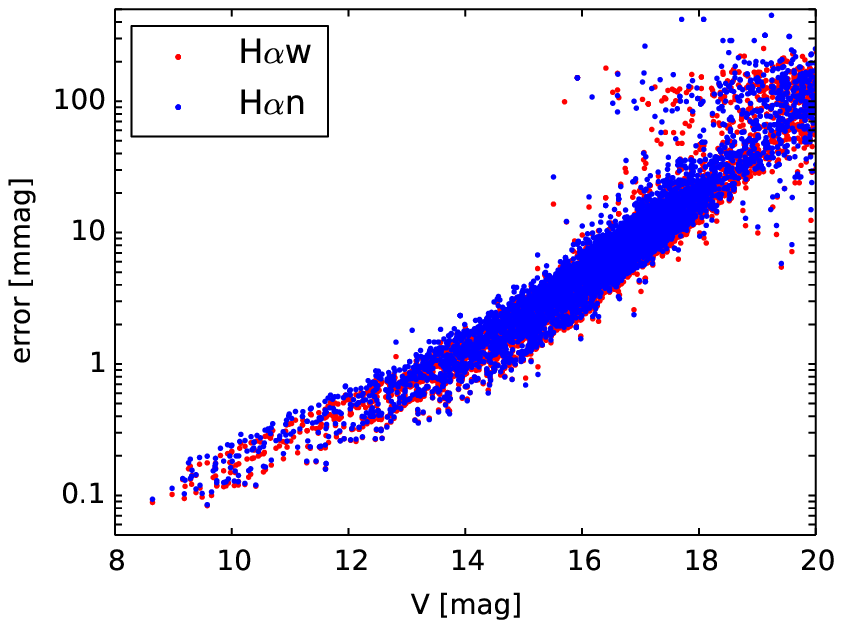}

{\large \bf b) \Halpha\ activity}\\

\includegraphics[angle=0,width=86mm,clip]{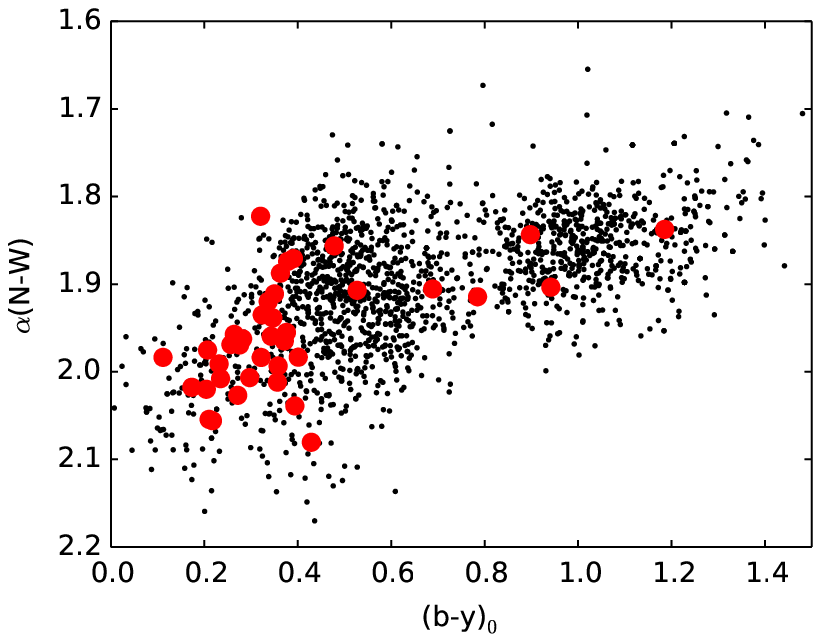}
\caption[ ]{\Halpha\ photometry of IC\,4756. {\bf a.} Standard errors in the STELLA \Halpha\ CCD data for both filters (n narrow, w wide). {\bf b.} \Halpha\ index $\alpha$ versus de-reddened $b-y$ color; an activity analog of the H-R diagram. \Halpha\ emission increases upward, \Halpha\ absorption increases downward. The thick dots indicate the CoRoT stars from Table~\ref{T1}. Spectral type and color equivalents are dF0, $b-y$=0\fm2; dK0 $b-y$=0\fm5; dM, $b-y$$>$0\fm8.
 \label{F6}}
\end{figure}

Fig.~\ref{F6}b shows the $\alpha$ index as defined in Eq.~\ref{eq1} as a function of $b-y$. This diagram is an activity analog of the H-R diagram because most of the targets form a main sequence which itself is split into a convection-dominated and a radiation-dominated part. The latter is characterized by the change of the strength of the \Halpha\ line with effective temperature and whether its formation is photo-ionization dominated or collision dominated. Brighter $\alpha$ index in Fig.~\ref{F6}b means more emission while fainter $\alpha$ index means more absorption.  Stars with pure (chromospheric) emission lines will lie above the main sequence, e.g., the active M dwarf sample of Herbst \& Layden~(\cite{herbst}).

The homogeneity of the points in Fig.~\ref{F6}b indicates that IC\,4756 is a magnetically quiet cluster containing no significantly active stars, except perhaps in close binaries. As such it resembles the Hyades. However, IC\,4756 has $\approx$10 giants whereas the Hyades is known to have just four which is in agreement with IC\,4756 being slightly older than the Hyades. The sample of CoRoT stars in this paper identifies only one cluster giant which is a spectroscopic binary with a period of 6.8~yrs (HSS\,314; Table~\ref{T5}). See also our individual notes.

\subsection{Equivalent widths and index calibration}

%Fig.~\ref{F-Haspec} shows representative \Halpha\ spectra for the double-lined binaries.
Because all of our CoRoT targets have at least one high-resolution \'echelle spectrum, we also attempt to verify the photometric $\alpha$ index with a spectroscopically measured equivalent width (EW). The \Halpha\ index is a dimensionless and non-physical quantity, and is thus not directly comparable to spectroscopic studies that usually measure an equivalent width in m\AA\ or a relative surface flux in erg\,cm$^{-2}$s$^{-1}$. The only shortcoming with our calibration of $\alpha$ vs. EW is the low S/N of most of the spectra. We also expect some intrinsic \Halpha\ variability for some of these targets but this effect is expected small compared to the measuring errors, in particular at the given cluster age.

\Halpha\ equivalent width is defined as the flux below the continuum absorbed by the full \Halpha\ line profile. The targets in Table~\ref{T1} exhibit rather different profiles ranging from wing-dominated hot dwarfs to wing-less cool giants. The latter, though, are mostly cluster non-members.  We employ the {\sl splot} routine in IRAF for the measurement and use a Lorentzian profile fit. Repeated measurements of one and the same spectrum allow an estimate of the internal error. It is dominated by the low S/N ratio and only to a much lesser extent by the fitting function and procedure. Repeated measurements when a target has more than just a single exposure allow an estimate of the external error. In our spectra the external error is driven by the uncertainty of the continuum level and is typically 10--20\%. For the faintest targets it may amount to 30\%\ (after all, these are $R$=55,000 spectra of 13th-mag stars with a 1.2m telescope).

\begin{figure}[!tbh]
\includegraphics[angle=0,width=86mm,clip]{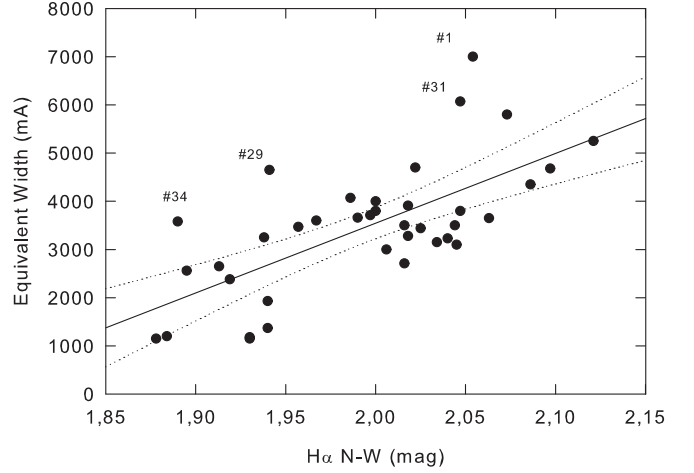}
\caption[ ]{Photometric \Halpha\ index N-W (narrow minus wide) in magnitudes vs. spectroscopic \Halpha\ equivalent width in milli Angstroem. The line is a simple regression fit and the dotted lines are its 95\%\ confidence levels. Four outliers are identified and discussed in the text.
 \label{F-cal}}
\end{figure}

Fig.~\ref{F-cal} shows the relation between the photometric $\alpha$ index and the spectroscopic equivalent width. A linear regression appears to be an adequate fit and is
\begin{equation}\label{eq2}
EW (m\AA) = -25418 + 14483 \ \alpha (N-W)\ .
\end{equation}
The standard deviation is 680\,m\AA . Four stars lie significantly above the bulk (\#1, 29, 31, 34). Star \#1 is a very broad-lined A star with an uncertain continuum setting, \#29 has very narrow lines but very extended \Halpha\ wings, \#31 is a SB2, and \#34 is our extreme flare star. If removed from the sample the reduced standard deviation is close to 550\,m\AA . There is a color dependency of \Halpha\ EW with $b-y$ which peaks at early A stars ($b-y\approx0\fm0$) and contributes to the residuals of the fit from Eq.~\ref{eq2}. A linear regression with Str\"omgren $b-y$ from the same sample of stars gives the functionality
\begin{equation}\label{eq3}
EW (m\AA) = 5781 - 4264 \ (b-y)_0 \ .
\end{equation}

%------------------------------   F7:  Ha vs. EW calibration

\subsection{Lithium abundances}

A total of 12 cluster members show a Li\,{\sc i} $\lambda$670.78-nm absorption line. Eleven are early-to-mid G dwarfs, one is a late-G or early-K dwarf. Lithium on the surface of cool main-sequence stars is usually interpreted to be a sign of youth and/or an enhanced mixing process possibly because of the presence of rotation and a magnetic field or the engulfing of a Li-rich large planet. We have two targets in our sample (\#36,\#37) in common with the three targets measured by Pace et al. (\cite{pace10}). Their spectroscopically determined effective temperatures and metallicities differ to ours by up to 300\,K and 0.1\,dex, respectively, while the lithium abundances are in reasonable agreement. We emphasize that our spectra are of comparatively low signal-to-noise ratio.

Lithium abundances in this paper are determined from the equivalent width and the effective temperature from the spectrum synthesis with the aid of the NLTE conversions from Pavlenko \& Magazzu (\cite{pm96}). Spectra have S/N$\approx$20--50:1 but multiple spectra allowed a repeated detection which was used to verify the Li\,{\sc i}-line. The nearby Ca\,{\sc i} 671.7-nm line was measured for comparison and its numerical values are given in the individual notes. A Gaussian fit or, if the S/N was very low, a simple line integration yielded the individual equivalent widths for the combined lithium isotopes. The equivalent width of the Fe\,{\sc i} 670.74 blend was taken from the NSO solar atlas (Wallace et al. \cite{nso}) and subtracted from the measurements. It is given in the individual notes. The time averaged Li results are listed in Table~\ref{T6}. Abundances are with respect to the $\log n(H)$=12.00 scale for hydrogen.

The IC\,4756 G-dwarfs have an average NLTE Li abundance of $\log n$=2.39$\pm$0.17(rms). The Li abundance for the SB2-primary HSS\,174a (\#\,17) is just a lower limit because the errors in the temperature (and the gravity) are very large (800\,K, 0.5\,dex). Another target (HSS\,95=\#16) shows Li\,{\sc i}-670.8 at the 50\,m\AA\ level but its $T_{\rm eff}$ determination is not reliable (see individual notes).

% --------------------------------- Table 6   Lithium stars
\begin{table}[!tbh]
\begin{flushleft}
\caption{Cluster members with lithium detection.}\label{T6}
\begin{tabular}{lllllll}
\hline \noalign{\smallskip}
No. & HSS & $T_\mathrm{eff}$ & $\log g$ &[Fe/H]     & $EW_{\rm Li}$   & $\log n$(Li)  \\
    &     & (K)              & (cm/s$^2$) & ($\odot$) & (m\AA) & (H=12) \\
\noalign{\smallskip}\hline \noalign{\smallskip}
8   &107a & 5700: & 4: & \dots          & 80 & 2.5: \\
    &107b & 5700: & 4: & \dots          & 80 & 2.5: \\
17  &174a & 5800: & 4.6: &\dots         & 45 & 2.5:  \\
26  &259  & 5200: & 4.4 &--0.9$\pm$0.5  & 98 & 2.2:\\
27  &284  & 5700 & 4.6 &--0.23$\pm$0.12 & 46 & 2.25$\pm$0.1  \\
29  &211  & 5680 & 4.6 &--0.20$\pm$0.06 & 30 & 2.1$\pm$0.1 \\
30  &335  & 5760 & 4.6 &+0.05$\pm$0.13 & 37 & 2.2$\pm$0.1  \\
31  &150a & 5670 & 4.5 &--0.12$\pm$0.09& 85 & 2.6$\pm$0.2  \\
    &150b & 5670 & 4.5 &\dots          & 76 & 2.5$\pm$0.2 \\
34  &74   & 5500 & 4.4 &--0.18$\pm$0.10& 64 & 2.2$\pm$0.1 \\
35  &222  & 5770 & 4.5 &--0.12$\pm$0.04& 66 & 2.45$\pm$0.1 \\
36  &240  & 5760 & 4.6 &--0.10$\pm$0.04& 84 & 2.6$\pm$0.1 \\
37  &165  & 5790 & 4.6 &--0.03$\pm$0.04& 83 & 2.6$\pm$0.1 \\
38  &\dots& 5500 & 4.6 &--0.25$\pm$0.10& 74 & 2.3$\pm$0.1 \\
\noalign{\smallskip}\hline
\end{tabular}
\end{flushleft}
\end{table}

\section{Summary and conclusions}\label{S7}

We have reported deep wide-field photometry of the open cluster IC\,4756, together with time-series photometric and spectroscopic monitoring of 38 brighter stars in the region of the cluster. IC\,4756 is particularly interesting because it is comparable in age with the Hyades. We have revised the cluster age to 890$\pm$70\,Myr and have determined an average reddening $E(b-y)$ of 0\fm16, a distance modulus of 8\fm02, and an average metallicity of --0.08$\pm$0.06. At this age of IC\,4756, we do not expect rapid stellar rotation anymore and thus no (or only marginal) magnetic activity. This is confirmed by our \Halpha\ photometry and spectroscopy, which also led to a calibration of the photometric \Halpha\ index with spectroscopically determined equivalent widths.

CoRoT photometry of 37 of the 38 targets enabled us to determine precise periods for 27 of them. Five stars with periods less than a day are all non-radial pulsators with a rich and complex frequency spectrum. The shortest rotation period is 1.68\,d for the late-F star HSS\,95 and the longest is 11.4\,d for the G2-dwarf HSS\,240. For six of the rotating stars, we see a splitting or asymmetric broadening of the main frequency and use this to determine surface differential rotation. Values of $\Delta\Omega/\Omega$ between 0.04 and 0.15\,rad\,d$^{-1}$ show them to have generally weaker differential rotation than the Sun.  Eleven G-stars were found with an average logarithmic lithium abundance of 2.41$\pm$0.17 (rms), almost identical to the G dwarfs in the Hyades (see Takeda et al. \cite{takeda}). Two of these targets are double-lined spectroscopic binaries where both G-components show a lithium line. In addition, ten new spectroscopic binaries were found in this paper and orbits have been presented for five of them. In a forthcoming paper, we will extend our CoRoT sample for IC\,4756, and aim to construct a color-period diagram.

\acknowledgements{We thank Annie Baglin and the CoRoT staff for their support with the present observations. STELLA was made possible by funding through the State of Brandenburg (MWFK) and the German Federal Ministry of Education and Research (BMBF). The facility is run in a collaborative way with the IAC in Tenerife, Spain, and we thank their mountain staff for their continuous support. Support by the DFG through grant STR 645/7-1 allowed us to initiate the STELLA Open Cluster Survey (SOCS), and obtain the data in the present paper. We also thank Robert Schwarz for helping with the WiFSIP data reduction and Federico Spada for discussions regarding the isochrones. An anonymous referee is warmly thanked for his/her careful reading which made this paper overall a better one.}

\appendix

\section{STELLA/WiFSIP observing log, transformation functions, and combined CCD images}

\begin{figure*}[!tbh]
\includegraphics[angle=0,width=18cm]{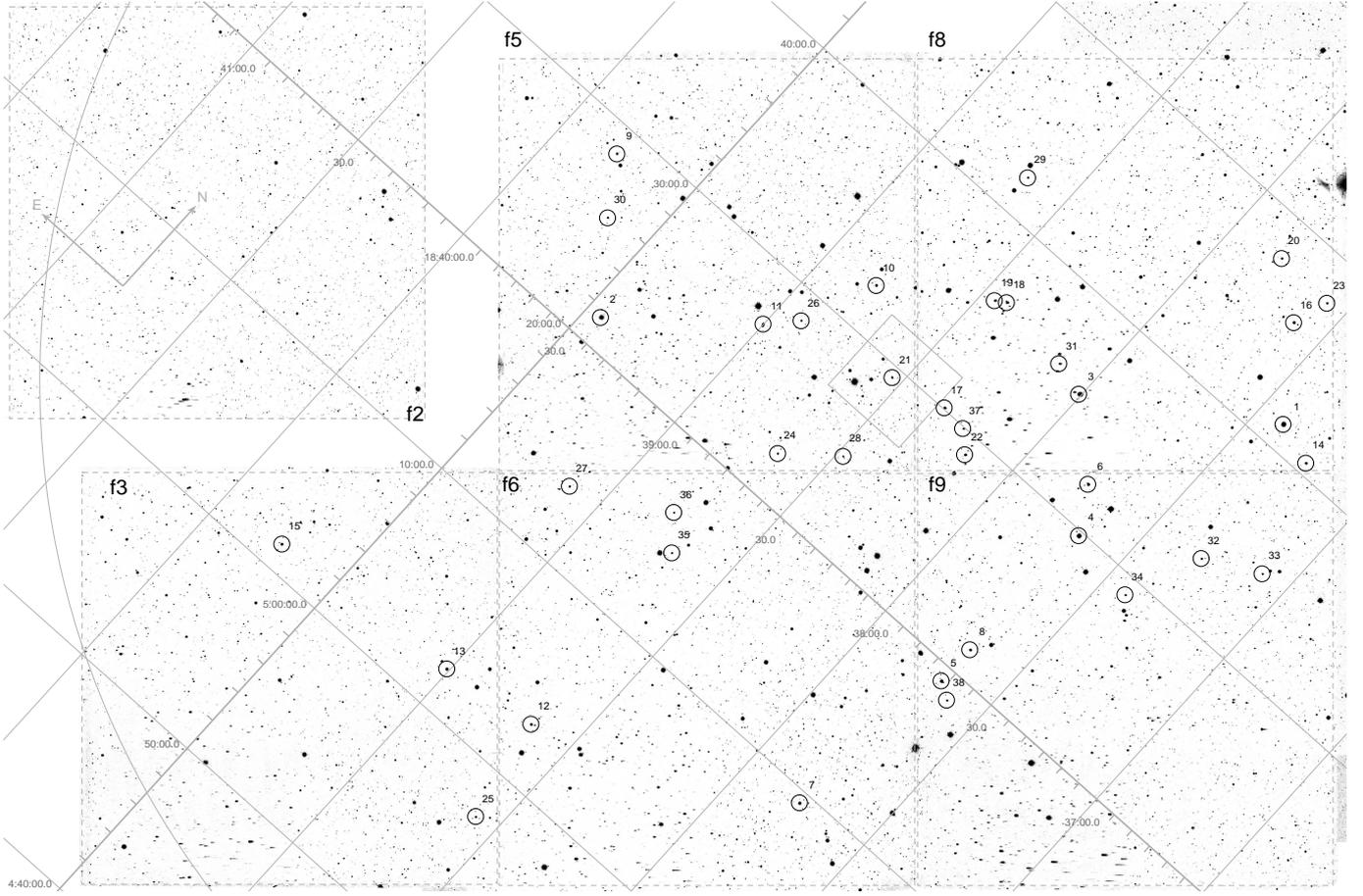}
\caption[ ]{STELLA/WiFSIP fields for IC~4756. Six fields were observed, indicated by the dashed boxes. Field f2 in the top left corner had been offset due to an intervening bright star. Each field is 22\arcmin$\times$22\arcmin\ and the total coverage is 0.67 square degrees or 38\%\ of the cluster. The approximate size of the cluster (90\arcmin ) is indicated by the large circle on the left. The cluster center is at 18h38m31.2s, +05\degr29\arcmin24\arcsec\ (2000.0) and is highlighted by the rotated square box. The stellar ID numbers are the running numbers from Table~\ref{T1} and identify the stars observed with CoRoT. The image is a $uby$ composite.
\label{Fapp1}}
\end{figure*}

Table~\ref{AT1} is the observing log of the Str\"omgren photometry with the Wide Field STELLA Imaging Photometer (WiFSIP). Only observing blocks are identified, not the individual CCD frames. One observing block consists of five consecutive integrations in $u$, $v$, $b$, $y$, H$\alpha$n, H$\alpha$w, H$\beta$n, and H$\beta$w. The table lists the mean J.D. given for the mid exposure, the field identification from Fig.~\ref{Fapp1}, the total number of CCD frames, total exposure time, the FWHM of the average stellar image, and the average airmass of the block.

The transformation equations are based on the calibration of Crawford \& Barnes (\cite{cra:bar}) and we followed the recipe from Balog et al. (\cite{balog}). Its four zero-point constants refer to the slopes for $(b-y)$, $m_1$, $c_1$ and $\beta$, respectively, while the three color-term coefficients are for $V$, $m_1$ and $c_1$. The WiFSIP field of view contained eight standard stars from the compilation of Hauck \& Mermilliod (\cite{hauck}) and six of them could be used to determine the zero points. However, all of these stars are brighter than 9th magnitude in $V$ and are thus close to the saturation limit of our CCD. Therefore, the standard deviations between the literature values and the WiFSIP instrumental values are comparably large, $\pm0.012$ for $(b-y)$, $\pm$0.07 for $c_1$, $\pm$0.04 for $m_1$ and $\pm$0.19 for $\beta$, and are not representative for the photometric precision. The $(b-y)$ color term in the equations for $m_1$ and $c_1$ was too uncertain and was set to zero.
\begin{eqnarray}
V = 0.90 + y_\mathrm{obs} - 0.14\cdot(b-y)\\
b-y = 1.10 + 0.88\cdot(b-y)_\mathrm{obs}\\
m1 = -0.08 + 0.78\cdot m1_\mathrm{obs}\\
c1 = -0.06 + 0.89\cdot c1_\mathrm{obs}\\
u-b = 1.90 + 0.92\cdot (u-b)_\mathrm{obs}\\
\beta = 2.65 + 1.80\cdot \beta_\mathrm{obs}\\
\alpha = 1.94 + \alpha_\mathrm{obs} \ .
\end{eqnarray}

The final WiFSIP Str\"omgren and \Halpha\ indices for the CoRoT stars are listed in Table~\ref{AT2}.

The epoch stacks were astrometrically re-calibrated in the tangent plane
projection with distortion polynomials using sources detected in Johnson $H$
or $K_{\rm s}$ from the 2MASS catalog (Cutri et al. \cite{cutri}), and those
with Johnson $J<17\fm5$.
An IRAF script iteratively invokes CCFIND, which correlates the catalog sources
with those in the image, and CCMAP, which computes the astrometric solution.
The Legendre fit to the projection transformation was cubic, with a threshold
of 2.4\,$\sigma$ to reject outlier astrometric points. The resulting projection
transformation had a standard deviation typically of $\la$0.10\arcsec\ in both
right ascension and declination.

The combined images for each bandpass represent very deep integrations with an
equivalent exposure time of, e.g., 3,900\,s in $y$ and up to 10,740\,s in
narrow-band H$\alpha$. Thereby, the IRAF FINDSTAR routine identified the
following numbers of stars per CCD field in the $y$ band:
f2 29,102;
f3 16,152;
f5 16,302;
f6 19,607;
f8 22,870; and
in f9 18,758 stars. The 5$\sigma$ limiting magnitude in $y$ is 21~mag.
Aperture photometry was then performed with the IRAF {\sl daophot} routine with
apertures adapted to the FWHM of the frame under the requirement of optimal
sampling. Average internal errors are basically photon-noise limited in all
bandpasses and reach 1\,mmag for stars between 12\fm5--13\fm5 and 0\fm2 at the
limiting magnitude in $y$. These rms values are obtained relative to a sky
background in the immediate vicinity of the numerical aperture.
External errors are estimated from the rms of all combined nightly frames.

%------------------------------ Table AT1:  WiFSIP obs log
\begin{table}
\begin{flushleft}
\caption{STELLA/WiFSIP observing log.}\label{AT1}
\begin{tabular}{llllll}
\hline \noalign{\smallskip}
MJD    & Field & \# of  & Tot. exp. & $\langle$FWHM$\rangle$ & Airmass \\
(245+) & I.D.  & frames & (sec)      & (arcsec) & \\
\hline \noalign{\smallskip}
5678.6010    & f8 &    39 & 2640 &   2.14  &  1.509 \\
5678.6616    & f8 &    39 & 2640 &   2.16  &  1.176 \\
5678.7234    & f8 &     4 &  365 &   2.47  &  1.082 \\
5685.5781    & f3 &    40 & 2800 &    2.10 &   1.566 \\
5688.5541    & f8 &     4 &  365 &   2.23  &  1.701 \\
5688.6636    & f8 &    40 & 2800 &    1.48 &   1.111 \\
5689.6219    & f9 &    40 & 2800 &    1.71 &   1.208 \\
5709.5474    & f9 &    40 & 2800 &    2.36 &   1.294 \\
5710.6358    & f3 &    40 & 2800 &    2.18 &   1.092 \\
5722.5374    & f6 &     2 &  182 &   2.63  &  1.188  \\
5723.4867    & f8 &    40 & 2800 &    2.00 &   1.438 \\
5723.6083    & f8 &    40 & 2800 &    1.88 &   1.086 \\
5725.5522    & f3 &    40 & 2800 &    2.16 &   1.137 \\
5733.4518    & f8 &    40 & 2800 &    1.86 &   1.501 \\
5734.6934    & f9 &    26 & 1304 &   1.40  &  1.476  \\
5737.6753    & f6 &    40 & 2800 &    1.67 &   1.408 \\
5738.5002    & f8 &    40 & 2800 &    1.97 &   1.169 \\
5738.5830    & f8 &    27 & 1340 &   2.59  &  1.092  \\
5740.5897    & f3 &    40 & 2800 &    1.37 &   1.114 \\
5741.4516    & f5 &    40 & 2800 &    1.67 &   1.350 \\
5741.6578    & f5 &    40 & 2800 &    1.53 &   1.355 \\
5742.5396    & f6 &    40 & 2800 &    1.62 &   1.092 \\
5743.4879    & f8 &    40 & 2800 &    2.39 &   1.165 \\
5743.5796    & f8 &    40 & 2800 &    3.18 &   1.106 \\
5743.6708    & f8 &    40 & 2800 &    2.27 &   1.494 \\
5744.4515    & f9 &    40 & 2800 &    1.36 &   1.297 \\
5745.5426    & f3 &    40 & 2800 &    1.68 &   1.092 \\
5746.6423    & f5 &    40 & 2800 &    1.97 &   1.344 \\
5747.4830    & f6 &    40 & 2800 &    2.06 &   1.154 \\
5748.4871    & f8 &    40 & 2800 &    1.75 &   1.133 \\
5752.4546    & f6 &     4 &  365 &   3.25  &  1.191  \\
5752.5432    & f6 &    40 & 2800 &    2.12 &   1.098 \\
5752.6358    & f6 &    40 & 2800 &    2.76 &   1.419 \\
5753.4498    & f8 &    40 & 2800 &    2.57 &   1.198 \\
5753.5408    & f8 &    40 & 2800 &    2.19 &   1.094 \\
5770.4190    & f3 &    40 & 2800 &    2.39 &   1.161 \\
5770.4892    & f3 &    39 & 2640 &   1.93  &  1.096  \\
5771.4432    & f5 &    39 & 2770 &   1.40  &  1.105  \\
5772.4710    & f6 &    39 & 2640 &   2.43  &  1.090  \\
5773.6300    & f8 &    40 & 2800 &    2.07 &   2.050 \\
5774.5479    & f9 &    40 & 2800 &    2.48 &   1.258 \\
5776.4404    & f5 &    40 & 2800 &    1.32 &   1.094 \\
5776.5417    & f5 &    40 & 2800 &    1.49 &   1.248 \\
5777.4411    & f6 &    40 & 2800 &    1.73 &   1.094 \\
5777.5426    & f6 &    40 & 2800 &    1.55 &   1.270 \\
5778.4405    & f8 &    40 & 2800 &    1.48 &   1.088 \\
5778.5219    & f8 &     3 &  274 &   2.20  &  1.188  \\
5779.5329    & f9 &    40 & 2800 &    2.77 &   1.252 \\
6157.3795    & f2 &    20 & 1200 &    1.10 &   1.116 \\
6157.4041    & f2 &    20 & 1200 &    1.18 &   1.090 \\
6157.4286    & f2 &    20 & 1200 &    1.31 &   1.089 \\
6157.4531    & f2 &    20 & 1200 &    1.20 &   1.114 \\
6157.4776    & f2 &    20 & 1200 &    1.26 &   1.168 \\
6157.5022    & f2 &    20 & 1200 &    1.26 &   1.256 \\
6158.3786    & f2 &    20 & 1200 &    1.25 &   1.113 \\
6158.4033    & f2 &    20 & 1200 &    1.22 &   1.089 \\
6158.4278    & f2 &    20 & 1200 &    1.21 &   1.090 \\
6158.4526    & f2 &    20 & 1200 &    1.30 &   1.118 \\
6158.4887    & f2 &    20 & 1200 &    1.39 &   1.212 \\
6158.5132    & f2 &    20 & 1200 &    1.61 &   1.326 \\
6158.5487    & f2 &    20 & 1200 &    1.82 &   1.606 \\
\noalign{\smallskip}\hline
\end{tabular}

\vspace{1mm}Notes. Field coordinates in degrees are:
f2(RA,DEC)=(280.1710,+5.2061); f3=(279.8442,+4.9322); f5=(279.8170,+5.4500);
f6=(279.5706,+5.1775); f8=(279.5432,+5.6953); f9=(279.2968,+5.4227).
\end{flushleft}
\end{table}

% --------------------------- AT2 appendix table
\begin{table}
\begin{flushleft}
\caption{Results from STELLA/WiFSIP Str\"omgren photometry.}\label{AT2}
 \begin{tabular}{lllllll}
  \hline \noalign{\smallskip}
 No.   &  $V$ & $(b-y)_0$ & [m1] & [c1] & \Halpha & \Hbeta \\
       &  \multicolumn{6}{c}{(mag)} \\
  \noalign{\smallskip} \hline \noalign{\smallskip}
 1 & 9.655 & 0.151 & 0.066 & 1.001 & 2.058 & 2.729\\
 2 & 9.744 & 0.505 & 0.394 & 0.266 & 1.947 & 2.570\\
 3 & 10.045 & 0.188 & 0.051 & 0.974 & 2.094 & 2.771\\
 4 & 10.003 & 0.185 & 0.196 & 0.806 & 2.015 & 2.844\\
 5 & 11.218 & 0.196 & 0.215 & 0.539 & 2.096 & 2.812\\
 6 & 11.764 & 0.242 & 0.186 & 0.586 & 1.997 & 2.838\\
 7 & 11.491 & 0.213 & 0.206 & 0.771 & 2.048 & 2.703\\
 8 & 12.362 & 0.325 & 0.191 & 0.304 & 1.979 & 2.663\\
 9 & 11.603 & 1.163 & 0.827 & --0.236 & 1.877 & 2.592\\
10 & 12.032 & 0.210 & 0.229 & 0.572 & 2.031 & 2.795\\
11 & 12.049 & 0.250 & 0.193 & 0.618 & 2.067 & 2.715\\
12 & 12.336 & 0.407 & 0.153 & 0.955 & 2.120 & 2.871\\
13 & 11.680 & 0.090 & 0.337 & 0.422 & 2.024 & 2.754\\
14 & 12.269 & 0.254 & 0.099 & 0.533 & 2.009 & 2.701\\
15 & 12.172 & 0.260 & 0.192 & 0.379 & 2.002 & 2.746\\
16 & 12.081 & 0.182 & 0.084 & 0.705 & 2.060 & 2.778\\
17 & 12.558 & 0.334 & 0.130 & 0.322 & 2.052 & 2.667\\
18 & 12.568 & 0.275 & 0.090 & 0.532 & 2.047 & 2.703\\
19 & 12.239 & 0.763 & 0.380 & 0.253 & 1.954 & 2.649\\
20 & 12.725 & 0.667 & 0.325 & 0.340 & 1.945 & 2.619\\
21 & 12.934 & 0.302 & 0.249 & 0.297 & 1.975 & 2.653\\
22 & 12.389 & 0.372 & --0.011 & 1.147 & 2.079 & 2.817\\
23 & 12.554 & 0.920 & 0.411 & 0.234 & 1.943 & 2.721\\
24 & 12.407 & 0.876 & 0.635 & --0.076 & 1.884 & 2.576\\
25 & 12.925 & 0.300 & 0.236 & 0.283 & 2.024 & 2.662\\
26 & 13.030 & 0.321 & 0.216 & 0.357 & 1.999 & 2.636\\
27 & 12.778 & 0.298 & 0.246 & 0.232 & 1.862 & 2.617\\
28 & 13.304 & 0.354 & 0.225 & 0.399 & 1.995 & 2.665\\
29 & 13.294 & 0.328 & 0.088 & 0.407 & 1.951 & 2.667\\
30 & 13.131 & 0.315 & 0.225 & 0.356 & 1.960 & 2.654\\
31 & 13.182 & 0.379 & 0.098 & 0.399 & 2.023 & 2.647\\
32 & 13.234 & 0.235 & 0.184 & 0.792 & 2.009 & 2.771\\
33 & 13.478 & 0.341 & 0.136 & 0.517 & 1.928 & 2.588\\
34 & 13.588 & 0.456 & 0.178 & 0.277 & 1.896 & 2.555\\
35 & 13.604 & 0.369 & 0.238 & 0.282 & 1.911 & 2.573\\
36 & 13.596 & 0.356 & 0.280 & 0.221 & 1.914 & 2.590\\
37 & 13.640 & 0.349 & 0.128 & 0.324 & 2.005 & 2.695\\
38 & 13.562 & 0.336 & 0.231 & 0.200 & 2.033 & 2.663\\
\noalign{\smallskip}\hline
\end{tabular}
\end{flushleft}
\end{table}

%---------------   Fapp1:  WiFSIP field mosaic

\section{Notes on individual targets}\label{S5}

\subsection*{\object{HSS 73} = \#1}

The CoRoT photometry in imagette modus is overexposed and consequently the relative brightness amplitudes are rather uncertain and are only for orientation. The periodicity of 0.4\,d of the light variations should not be affected, but the amplitude is so close to the noise that we are uncertain about its reality and dismiss it. The value in Table~\ref{T3} is thus very tentative. The spectra are those of a mid A-star with very broad Balmer lines with equivalent widths up to 20\,\AA\ for \Halpha , 14\,\AA\ for \Hbeta , and 10\,m\AA\ for \Hgamma . Our radial velocities are only estimates. The diffuse interstellar band (DIB) lines at $\lambda$661.3, 578.0 and 579.7 are clearly detected and match those of other cluster members ($\lambda$862.1 is not detected).

\subsection*{\object{HSS 314} = \#2}

Thirtynine radial velocities confirm this star (also designated Kopff~139) to be a long-period SB1. An orbit was published earlier by Mermilliod et al. (\cite{mer:and}) with a period of 3834$\pm$36\,d (10.5\,yrs), an eccentricity of 0.22 and a $K$-amplitude of just 2.6\,\kms . Our data span a full amplitude of 4.4\,\kms\ already within 843\,d (2.3~yrs) and it is thus possible that above orbital period is wrong. However, our data are not sufficient in coverage yet to compute a definite revised orbit but a preliminary circular orbit from STELLA data alone is given in Table~\ref{T5} with $P_{\rm orb}$=2487\,d (6.8\,yrs). PARSES finds the best fit with $T_{\rm eff}$=5070$\pm$50\,K, $\log g$=3.8$\pm$0.1, $v\sin i$=3.9$\pm$0.3\,\kms , and [Fe/H]=--0.16$\pm$0.03.

\subsection*{\object{HSS 138} = \#3}

The Balmer lines appear broad with equivalent widths of 3.5\,\AA\ for \Halpha , 3.0\,\AA\ for \Hbeta , and 3.0\,m\AA\ for \Hgamma . The interstellar Na\,D lines are sharp and very prominent. Radial velocities are difficult owing to a lack of spectral lines. The wavelengths with respect to the DIB lines are in agreement with the expected cluster velocity and we are confident that the star is a member. A PARSES solution of the best spectrum suggests $T_{\rm eff}$=6900\,K, $\log g$=3.5, $v\sin i$=100\,\kms , and [Fe/H]=--0.45 which suggests either a late-A or a slightly evolved mid-F star.

\subsection*{\object{HSS 106} = \#4}

The star is a single-lined spectroscopic binary with an orbital period of 23\,d (Table~\ref{T5}). We sometimes see a double-lined spectrum due to a third star whose velocities appear constant. It is unclear whether this star is a background or confusing source or part of a long-period triple system. No clear photometric period could be extracted from the CoRoT photometry.

\subsection*{\object{HSS 108} = \#5}

Pulsating star in a single-lined spectroscopic binary. We have 40 RVs that suggest a low-amplitude of just 4\,\kms\ with an orbital period of 25.7$\pm$0.15\,d. The orbit is so uncertain that we want to mention it only in these notes (other elements are $T_0$=2,456,391, $\gamma$=--13.5$\pm$1\,\kms , $e$=0, $a_1\sin i$=1.42$\pm$0.29\,10$^6$\,km, and $f(M)$=0.0002$\pm$0.0001). The fundamental pulsation period is 0.438\,d.

\subsection*{\object{HSS 113} = \#6}

The Balmer lines appears very broad with equivalent widths of 4\,\AA\ for \Halpha , 2.9\,\AA\ for \Hbeta , and 3\,\AA\ for \Hgamma . The interstellar Na\,D lines are sharp and prominent. The target is an A-F star with $v\sin i$ of $\approx$250$\pm$50\,\kms . The star pulsates with a period of 0.353\,d which is likely the fundamental pulsation period in a Delta Scuti-type star. Therefore, radial velocities are very difficult to obtain and we judge the value of $-10\pm4$\,\kms\ to be still in overall agreement with the cluster mean.

\subsection*{\object{HSS 112} = \#7}

The STELLA radial velocities do not allow a membership statement. The target has broad Balmer lines suggestive of $v\sin i$ of $\approx$150$\pm$20\,\kms . However, no pulsation pattern like in HSS\,113 is seen.  The periods from the CoRoT light curve are ambiguous, a FFT gives a 14-d value while the ACF suggests $\approx$40~d.

\subsection*{\object{HSS 107} = \#8}

Double-lined spectroscopic binary of two nearly equal components. An orbit with $P_{\rm orb}$=11.65\,d is given in Table~\ref{T5}. Na\,D has a strong interstellar absorption component and is partly affected by geocoronal emission. Both components show a Li\,{\sc i} $\lambda$670.8-nm line with an equivalent width of 80$\pm$4~m\AA\ compared to 110$\pm$5\,m\AA\ for the nearby Ca\,{\sc i} 671.7 line (both values corrected for two equal continua). Line ratios in the 640-nm region suggest an early-G spectral type and convert the Li equivalent width into an abundance of $\log n\approx 2.5$ (NLTE) under the assumption of $T_{\rm eff,1}$=$T_{\rm eff,2}$=5700\,K. We measure $v\sin i$ for both components of 4.5$\pm$1.5\,\kms .

The CoRoT data gave uncertain but consistent photometric periods from a FFT (10$\pm$0.5~d) and from the ACF (11$\pm$0.5~d). The average is adopted in Table~\ref{T3}.

\subsection*{\object{HSS 356} = \#9}

Not a cluster member. The SES spectra show it a K5III giant with an average radial velocity of --19.3\,\kms , almost a perfect match with the M-K standard $\alpha$~Tau. The Balmer lines appear as narrow absorption lines without noticeable wings. Optically thin lines appear also sharp with only small rotational broadening of $v\sin i$=4.5$\pm$0.5\,\kms . No Ca\,{\sc ii} H\&K nor IRT emission nor Li\,{\sc i} 670.8-nm or He\,{\sc i} 587.5-nm absorption is detectable. The star is thus not chromospherically active. The red to near-IR part of the spectrum is dominated by molecular lines, e.g., the typical TiO bandheads, e.g., at 705.5\,nm. The Na\,D doublet appears very asymmetric with a broad wing only on the respective blue-wavelength sides but sharp on the red sides. This is due to a stationary emission component redshifted by +43\,\kms\ that is also seen at \Halpha\ and K\,{\sc i} 769.9\,nm where it appears with a redshift of $\approx$42\,\kms . The DIB line at $\lambda$661.3\,nm has an EW of 180\,m\AA . The forbidden oxygen lines from [O\,{\sc i}] at 557.7, 630.0, and 636.4~nm appear with line intensities relative to the continuum of up to 5, 0.5, and n.d., respectively, and are all mostly of geocoronal origin.

Our PARSES analysis suggests an effective temperature of 3890$\pm$50\,K and $\log g$ of 1.4$\pm$0.5. The large error in the metallicity, --0.01$\pm$0.22, indicates cross talk in the solution with $\log g$, and thus it is likely that both values are not real. The expected absolute magnitude of a K5 giant is --0\fm2$\pm$0.2 and its distance modulus is $\approx$11\fm7. Thus, the star is in the cluster background at $\approx$1.9\,kpc.

The CoRoT light curve clearly indicates changing variability with an amplitude of up 0\fm15 and a period of 4.25\,d. The FFT reveals three discrete peaks at 4.257\,d (main peak) 5.253\,d and 3.629\,d. The FWHM is 0.21\,d and the formal $\Delta\Omega/\Omega$ would be 0.782$\pm$0.003. The highly variable and modulated light curve and the three isolated periods along with a highly variable ACF could either indicate three different active longitudes or systematic spot evolution. When $P$ is interpreted as the rotational period of the star the minimum radius $R\sin i$ would only be 0.38\,R$_\odot$. Furthermore, if we assumed a radius typical for a K5 giant of 25\,R$_\odot$ the inclination would be below 1$^\circ$. We note that the K5III MK-standard \object{$\alpha$\,Tau} has a measured radius of 44\,$R_\odot$ (Richichi \& Roccatagliata \cite{ri:ro}). In any case, it is highly unlikely that the 4.25-d period is the rotation period of the K5 giant but could stem from a very close-by undetected and large-amplitude foreground star.

\subsection*{\object{HSS 241} = \#10}

Pulsating star with 0.8-d fundamental period. Six radial velocities, evenly distributed within two months in 2013, show a steady decrease from --11.4 to --25.6\,\kms . It suggests a SB1-binary with yet unknown systemic velocity. The radial velocity in Table~\ref{T1} is a current average. The DIB lines at $\lambda$661.3, 578.0, and 579.7 are of comparable strength than for cluster members.

\subsection*{\object{HSS 269} = \#11}

It appears that our STELLA SES data base of HSS\,269 consists of three different stars. One target was a plain misidentification (\object{HSS 279}; not a cluster member, see below). The other target is a very close neighbor of comparable brightness (\object{HSS 267}, separated by 6.7\arcsec ). STELLA acquired the correct target in normal weather conditions but the guider may have accidentally picked up the close target during occasions of high wind gusts during at least three occasions. These data were removed but the overall rms of our RVs of HSS\,269 remained larger than expected. In addition, CoRoT did not resolve the two targets and thus its photometry is always for both stars.

The STELLA spectrum 20130814B-0006 is for \object{HD 172248} (=HSS\,279) and shows a strong He\,{\sc i} $\lambda$587.5 line with an EW of 160\,m\AA . The two Si\,{\sc ii} lines at $\lambda$634.7 and $\lambda$637.1 are measured with EWs of 182\,m\AA\ and 150\,m\AA , respectively. The entire spectrum resembles that of a B supergiant. The neutral potassium line K\,{\sc i} 769.9\,nm is detected with an EW of 40\,m\AA\ and the O\,{\sc i}-777\,nm triplet is well resolved with a residual intensity of 0.75. However, no Ca\,{\sc ii} irt lines are apparent nor do we see the hydrogen Paschen series although a very broad feature spanning the entire echelle order coincides with P12 $\lambda$875\,nm. PARSES does not converge on a solution, which indicates parameters out of its range (which would be the case for a hot supergiant); $v\sin i$ appears unmeasurable but the radial velocity is --12.8\,\kms . The DIB lines at $\lambda$661.3, 578.0 and 579.7 are clearly detected with EWs of 98, 125, and 35\,m\AA\ and appear stronger than those of cluster members. We conclude that HSS\,279 is not a cluster member but a background early-B supergiant.

In comparison, the other two STELLA spectra obtained on April 8 and June 14, 2013 show a comparable but not identical spectrum to HSS\,279, in particular the He\,{\sc i} line is completely missing. The two Si\,{\sc ii} lines appear weaker with EWs of 100 and 55\,m\AA , respectively. The Ca\,{\sc ii} irt lines appear weak but are clearly detected, and again no sign of the hydrogen Paschen series. Radial velocities are --23.9 and --20.6\,\kms\ for the two spectra, respectively, which fit the cluster mean, and $v\sin i$ is $\approx$13\,\kms . The DIB lines are detected but are much weaker than in the spectrum of HSS\,279 and in agreement with the cluster distance. \Halpha\ is typical for a late-F dwarf in agreement with its $(b-y)_0$ of 0\fm316 from WiFSIP.

The combined CoRoT light curve of HSS\,269 and HSS\,267 shows a clear and modulated variability with a period of 2.0334\,d. If indeed a late-F dwarf, this period is likely its rotation period due to spots.

\subsection*{\object{HSS 208} = \#12}

Background star with strong DIB lines of 208, 420, 110\,m\AA\ at $\lambda$661.3, $\lambda$578.0 and  $\lambda$579.7, respectively. The Balmer and Paschen serie of hydrogen are dominating the spectrum (EW \Halpha\ 3.8\,\AA ; EW \Hbeta\ 4.2\,\AA ). The ISM Na\,D lines are saturated and the K\,{\sc i}-769.9 line has an EW of 130\,m\AA. The spectrum of this star is otherwise almost featureless. The CoRoT photometry shows an oscillation spectrum with a fundamental period of 0.154\,d.

\subsection*{\object{HSS 270} = \#13}

Pulsating hot star with very broad spectral lines. The O\,{\sc i} triplet is not resolvable. In contrast, K\,{\sc i}\,7699/7665 as well as the Na\,D lines appear very sharp. The Na\,D lines reach basically zero intensity in the core. If these lines are of ISM origin then the target must be a background star. Then, the average radial velocity of --22.2$\pm$2.5\,\kms\ must be a chance agreement with the cluster mean. Unfortunately, this part of the cluster is affected by intervening dust clouds as seen from the WISE images at 4.6$\mu$m and 12$\mu$m and thus the WiFSIP-based de-reddening may be inappropriate. Therefore, we cannot firmly conclude on its membership but removed the target from the isochrone fit.

\subsection*{\object{HSS 57} = \#14}

The CoRoT data reveal two periods of 10.7~d and $\approx$1~d that are both seen by naked eye in Fig.~\ref{F1}. Our  spectra indicate a He\,{\sc i} $\lambda$587.5 absorption line with an EW of 120\,m\AA . Both Na\,D lines are dominated by the sharp-lined ISM components.  Si\,{\sc ii} at $\lambda$634.7 and $\lambda$637.1 is measured with EWs of 120\,m\AA\ and 50\,m\AA , respectively. The He\,{\sc i} line and the two Si\,{\sc ii} lines indicate a hot star. The rest of the spectrum morphology suggests an early-G or late-F dwarf and the 10.7-d CoRoT period appears to be its rotation period. It is possible that the spectrum is a combination of two stars.
%XXX could this be two different stars? xxx

\subsection*{\object{HSS 348} = \#15}

Double-lined spectroscopic binary with an eclipsing tertiary star.

\subsection*{\object{HSS 95} = \#16}

The O\,{\sc i} triplet at 777\,nm appears blended but is clearly detected with a combined EW of 557\,m\AA . The traces of Li\,{\sc i} 670.7 are very weak due to the low S/N ratio but appears detected on an EW$\approx$50\,m\AA\ level. Other spectrum tracers, like the H$\beta$, H$\gamma$, Ca\,{\sc ii} irt and many of the strong Fe\,{\sc i} lines in the blue region, indicate a rapidly-rotating late-F star with $v\sin i$=31$\pm$4\,\kms\ in agreement with the 1.68-d CoRoT period. A strong redshifted ISM line with EW of 90\,m\AA\ is seen at K\,{\sc i}~769.9\,nm (and also at Na\,{\sc i} D). The DIB line at $\lambda$661.3 has an EW of $\approx$75\,m\AA .

The CoRoT light curve in Fig.~\ref{F1} clearly indicates period doubling due to two migrating starspots. The full  amplitude doubles from 5\,mmag to 10\,mmag from double-humped to single-humped light curves. The FFT has its main peak at 1.6754\,d and a separated peak at 1.7628\,d, another period may be real at 1.6096\,d but is within the FWHM of 0.0785\,d of the main peak.

\subsection*{\object{HSS 174} = \#17}

Double-lined spectroscopic binary where the weaker component is narrow lined and the stronger component broad lined. The spectrum synthesis picks up only the stronger component and suggests a late-type star with $T_{\rm eff}$=5000$\pm$800\,K, $\log g$=4.1$\pm$0.5, and a $v\sin i$ of 14$\pm$2\,\kms . However, its solution is affected by the secondary lines. The metallicity is unconstrained in our fits. $V\sin i$ of the secondary is $<4$\,\kms . We measure a Li\,{\sc i} 670.8-nm line from the strong-lined component with an equivalent width of 47$\pm$3~m\AA\ (compared to 60$\pm$4\,m\AA\ for the nearby Ca\,{\sc i} 671.7 line). The spectrum morphology and various line ratios in the 643-nm region suggest two early-G stars with effective temperatures and gravities more near the upper limits of the PARSES results. We assume 5800\,K and $\log g$=4.6 and a continuum ratio of 2:1 (primary to secondary) to convert above Li equivalent width into an abundance.

The FFT shows a single peak at 9.115\,d and a smaller peak at 12.14\,d. The ACF indicates periods at 9.357\,d and 8.2475\,d, with the lower period still within the FWHM of 1.86\,d of the main peak.

\subsection*{\object{HSS 185} = \#19}

The spectrum synthesis fits of five spectra consistently give $T_{\rm eff}$=4700$\pm$100\,K, $\log g$=3.6$\pm$0.3, [Fe/H]=+0.2$\pm$0.1, and a $v\sin i$ of 4$\pm$1\,\kms . Its many sharp photospheric lines, the wing-less Balmer lines, and several line-depth ratios at 643\,nm are in agreement with a K0-2\,III-IV luminosity classification. The star is therefore not a cluster member. No measurable Li $\lambda$670.78 line is present (EW$<$10\,m\AA).

\subsection*{\object{USNO-B1 0957-0380847} = \#20}

Not a cluster member. The PARSES effective temperature is 4,630$\pm$100\,K, $\log g$=1.5$\pm$0.3 and [Fe/H]=--0.45$\pm$0.1. $V\sin i$ is basically not measurable with a nominal value of 1$\pm$1\,\kms . The interstellar Na\,{\sc i}\,D lines appear very complex and very strong and are split into two broad complexes separated by 1.0\,\AA , very similar to star \#23. The DIB line at $\lambda$661.3 has an EW of around 100\,m\AA . The \Halpha\ profile shows no wings and appears typical for a giant. The star is in the background and is likely a K-type giant. No measurable Li $\lambda$670.78 line is present (EW$<$10\,m\AA).

\subsection*{\object{HSS 194} = \#21}

Both \Halpha\ and Na\,D show a red-shifted, stationary, and sharp emission-line component. Its redshift is 48\,\kms\ for both D lines and $\approx$48\,\kms\ for \Halpha. The D$_2$ emission appears much stronger (residual intensity of 1.4) than the D$_1$ emission (residual intensity of around 1.0). Na\,D appears basically without wings and shows subtle line-profile structure most likely due to intervening interstellar absorption. The forbidden oxygen lines from [O\,{\sc i}] at 557.7, 630.0, and 636.4~nm rise above the continuum  with relative line intensities of up to 12, 5, and 2, respectively. We believe that all these emission lines are of geo-coronal origin.

The ACF shows a different signal after about seven variability cycles, i.e., after $\approx 23$\,d, with two distinct maxima of 3.33375\,d and 3.51875\,d. The two peaks are also visible in the FFT (Fig.~\ref{F-diff} and we interpret them as due to two spots co-rotating at two different latitudes. The respective errors are estimated from the FWHM to be 0.13\,d.

\subsection*{\object{HSS 158} = \#22}

Probably not a cluster member. Close binary, likely an eclipsing $\beta$\,Lyrae-type system. Therefore, the period in Table~\ref{T3} is the orbital period (which is likely also the rotation period because one would expect a bound rotation). The spectrum appears to be that of an A star. Five RVs show a range of --24\,\kms\ to +21\,\kms . A single spectrum taken on Mar. 30, 2013 shows it with a sharp central hump in the \Halpha\ core, most likely due to the RV shift of the two stellar components. All spectra show the Paschen series until P12 at 875.0\,nm. Strong DIB lines, e.g., at 6613\,\AA , also indicate rather a background target than a cluster member.

\subsection*{\object{HSS 89} = \#23}

Not a cluster member. The PARSES effective temperature is 4,560$\pm$150\,K, $\log g$=3.3$\pm$0.1 and [Fe/H]=--0.02$\pm$0.02. $V\sin i$ is basically not measurable with an upper limit of 2\,\kms . The interstellar Na\,{\sc i}\,D lines appear very complex and very strong and are split into two broad complexes.  The DIB line at $\lambda$661.3 has an EW of around 120\,m\AA , but difficult to measure. The star is a background early-K giant. No measurable Li $\lambda$670.78 line is present (EW$<$10\,m\AA).

\subsection*{\object{HSS 198} = \#24}

Not a cluster member. The spectrum-synthesis fits of five spectra consistently give $T_{\rm eff}$=4,570$\pm$110\,K, $\log g$=2.5$\pm$1.0, [Fe/H]=--0.2$\pm$0.3, and a $v\sin i$ of 2$\pm$1\,\kms . The star seems to be a background K2 giant {\rm or subgiant with strong and complex Na\,D and K\,{\sc i} line profiles}. No Li $\lambda$670.78 feature greater than 10\,m\AA\ is detectable. The 7-d period from the ACF could be an alias of a longer period. The Lomb-Scargle periodogram gave 7.215$\pm$0.03\,d with a smaller amplitude than ACF (4 instead of 6\,mmag) but an almost equally strong second period at around 16\,d.

\subsection*{\object{HSS 209} = \#25}

Our PARSES converged on three of the five spectra with average values of $T_{\rm eff}$=6060$\pm$240\,K, $\log g$=4.7$\pm$0.2, [Fe/H]=--0.2$\pm$0.2, and a $v\sin i$ of 22$\pm$4\,\kms . These parameters suggests an $\approx$F9 dwarf star. Five radial velocities outline a weak linear trend from --24.6\,\kms\ in early May 2013 to --23.3\,\kms\ by the end of August 2013 with an unweighted average of --23.9$\pm$0.6\,\kms\ (rms). The star could be a long-period binary but the RV evidence is not yet convincing. Strong red-shifted ISM lines at Na\,D and K\,{\sc i} 769.9\,nm dominate its profiles. The 3-day period in Table~\ref{T3} is from a Lomb-Scargle analysis and is very uncertain, the ACF did not give a convincing period at all. Three days is nevertheless in agreement with the measured $v\sin i$ and the expected radius of an F9 dwarfs and is thus most likely the rotation period of the star.

\subsection*{\object{HSS 259} = \#26}

Five spectra indicate average values of $T_{\rm eff}$=5200$\pm$400\,K, $\log g$=4.35$\pm$0.25, [Fe/H]=--0.9$\pm$0.5, and a $v\sin i$ of 10$\pm$2\,\kms . The star is a rapidly rotating late-G or early-K dwarf. A strong Li\,{\sc i}-670.8 line is detected in all of the spectra with an equivalent width of 104$\pm$6~m\AA\ (compared to 100$\pm$6\,m\AA\ for the nearby Ca\,{\sc i} $\lambda$671.7 line). We subtracted an equivalent width of 6\,m\AA\ from above value to obtain the logarithmic abundance of $\log n\approx$2.2 in Table~\ref{T6}. The light curve shows occasionally repeated dips that indicate a 5.9-day period, but overall it is not well defined and must be judged preliminary. However, a 5.9-d period is in fair agreement with the $v\sin i$ and the radius of a late-G dwarf.

\subsection*{\object{HSS 284} = \#27}

The ACF indicates two close periods at 3.8051\,d and 3.8688\,d. The FFT periodogram shows only one peak at 3.867\,d with a FWHM of 0.3\,d. The LSSD revealed two comparable periods of 3.874\,d and 4.346\,d. Four radial velocities show a linear change from --22.2 to --24.6\,\kms\ within two months in 2013 indicative of a long-period SB1-binary. The spectrum morphology and two PARSES solutions suggest an effective temperature of 5700$\pm$200\,K, $\log g$ of 4.6$\pm$0.1, [Fe/H]=--0.23$\pm$0.2 and $v\sin i$=13$\pm$3\,\kms . The star is a rapidly rotating early-G dwarf. A Li\,{\sc i}-670.8 line is detected in one of the spectra with an equivalent width of 46$\pm$3~m\AA\ (compared to 81$\pm$4\,m\AA\ for the nearby Ca\,{\sc i} $\lambda$671.7 line).

\subsection*{\object{HSS 189} = \#28}

Double-lined spectroscopic binary. Table~\ref{T5} gives a preliminary orbit based on 8 respective 6 measurements. The period of 110\,d is accordingly uncertain. No Li\,{\sc i}-670.8 lines detected. The spectrum morphology around the $\lambda$643-nm region suggests two mid F-dwarf stars. The ACF period of 7.2\,d might be spurious. A Lomb-Scargle periodogram gives three equally strong peaks at 12.15\,d, 6.73\,d and 5.27\,d but with smaller amplitudes than the ACF period.

\subsection*{\object{HSS 211} = \#29}

An early G-dwarf star. PARSES solutions for five spectra suggest $T_{\rm eff}=5,680\pm170$~K, $\log g$=4.63$\pm$0.03, and [Fe/H]=$-0.20\pm0.06$ with $v\sin i$ of 9$\pm$2\,\kms . A Li\,{\sc i}-670.8 line is detected in four of the seven spectra with an average equivalent width of 30$\pm$3~m\AA\ (compared to 100$\pm$4\,m\AA\ for the nearby Ca\,{\sc i} 671.7 line).

\subsection*{\object{HSS 335} = \#30}

The target was originally misidentified in our STELLA SES observations and was accidentally observed 26 times. For its radial velocities, we had chosen a synthetic template spectrum with 6,250\,K, $\log g$=4.0 and [Fe/H]=--0.5. The velocities appear to be constant near the cluster mean.

The $UBV$ magnitudes from Herzog et al. (\cite{herz}) are $B$=13\fm59, $V$=12\fm92, and $B-V$=0.67 and $U-B$=0.11 while NOMAD lists magnitudes in $BVR$ of 13\fm46, 13\fm28, and 12\fm52~mag. Herzog et al. (\cite{herz}) assigned a 90\%\ probability that it is a cluster member. Assuming a 0.8 mag/kpc absorption in the cluster direction, we estimate an absolute magnitude of +4\fm6 and a $(B-V)_0$ of 0\fm57, suggestive of an $\approx$G0 dwarf star. HSS\,335 was not detected in the ROSAT HRI survey by Randich et al. (\cite{ran}) and thus constrains its X-ray luminosity to less then $\approx$2\,10$^{29}$ erg\,s$^{-1}$.

The PARSES solution yields $T_{\rm eff}=5760\pm110$~K, $\log g$=4.61$\pm$0.05, and [Fe/H]=$+0.05\pm0.13$. The rotational broadening is $v\sin i$=9$\pm$1~\kms\ with an adopted macroturbulence of 4~\kms . The metallicity of HSS\,335 is actually consistent with the large range of cluster metallicities in the literature (see Sect.~\ref{S3.1}). The effective temperature and the gravity suggest a solar like (G2) star. Various Fe\,{\sc ii} lines are detected. In particular the ratio of the Fe\,{\sc i}-643.0 to the Fe\,{\sc ii}-643.2 line and the ratio of the Fe\,{\sc ii}-645.7 to the Ca\,{\sc i} 645.6-line indicate a late-F to early-G spectral class. The luminosity indicator Sr\,{\sc ii} 407.7~nm is detectable with an equivalent width of 150~m\AA\ and appears with half of the strength compared to the nearby Fe\,{\sc i} lines 407.1 and 406.3~nm. Its equivalent-width measure is rather uncertain owing to the low S/N but the profile appears unsaturated and thus more compatible with a dwarf luminosity classification than with a giant.

Profiles from Balmer \Halpha , H$\beta$, and H$\gamma$ appear broad winged with a narrow absorption core, typical for dwarf stars. Their equivalent widths are 5.20~\AA , 4.86~\AA , and 2.60~\AA , respectively; for example, for \Halpha , 1.3~\AA\ are contained in the core, the rest in the wings. The Ca\,{\sc ii} IRT profiles also show a comparable profile with a central absorption core. No Ca\,{\sc ii} H\&K emission lines are evident. Furthermore, no He\,{\sc i}~D3 feature is detectable and we can conclude that HSS\,335 is at least not a magnetically active star.

We find absorption from the Li\,{\sc i} 670.78-nm resonance line. Its measured equivalent width is 43$\pm$5~m\AA \ (approximately half of the nearby Ca\,{\sc i} 671.7-nm line which has 90$\pm$7~m\AA ). The Li line is still blended with a close-by, very weak, Fe\,{\sc i} line that cannot be measured in our spectra. Its equivalent width in the Sun (G2V) is 6~m\AA\ while it is 9~m\AA\ in a well-exposed spectrum of $\epsilon$~Eri (K0V). We assume it to be 6~m\AA\ for HSS\,335 (G0-2) and subtract it from the equivalent-width measure. This translates to a lithium abundance of $\log n(Li)=2.2\pm0.1$, based on the usual $\log n=12$ for hydrogen and an effective temperature of 5760~K.

By far the strongest emission lines are the forbidden oxygen lines from [O\,{\sc i}] at 557.7, 630.0, and 636.4~nm with line intensities relative to the continuum of up to 15, 20, and 8, respectively. The line intensities vary by many factors from night to night accompanied by variations of the radial velocity. The changes are perfectly consistent for the three line. No geocoronal emission is seen within the Na\,{\sc i} D absorption lines though (like in, e.g., HSS\,194, 356, and others), suggesting that the [O\,{\sc i}] emission could be partly of interstellar origin. However, we do not see any signs of the low-density sensitive [S\,{\sc ii}] or [N\,{\sc ii}] emissions at 671.6nm, 673.1nm and 654.8nm, 658.4nm, respectively, which again suggests that the [O\,{\sc i}] emission is of geocoronal origin. Numerous weak and narrow emission lines appear partially with a double-peaked profile, mostly at wavelengths red-ward of 700~nm. The average width at continuum level of these lines is $\approx$15~\kms\ (one half of the rotational broadening of the absorption lines). Following Hanuschik (\cite{hanu}), these lines are of telluric origin due to Earth's airglow layer.

Both photospheric Na\,{\sc i}~D lines are flanked on the blue side by an additional absorption line of comparable width. The displacement is --15.6~\kms\ from the cores of the photospheric D lines and their equivalent width is approximately one half of that of the stellar lines. It is generally accepted to be of interstellar origin and related to the column density of sodium in interstellar clouds along the photon path. Hobbs (\cite{hobbs}) obtained an empiric distance calibration of the interstellar D$_2$ equivalent width. If the equivalent width of $W_{\rm D2}$=180$\pm$10~m\AA\ of HSS\,335 is indeed all due to interstellar absorption, then the Hobbs-relation suggests a distance of $\approx$300~pc. This might still be in agreement with the revised \emph{Hipparcos} cluster distance of 440\,pc.

\subsection*{\object{HSS 150} = \#31}

Double-lined spectroscopic binary with two comparable components. A synthetic spectrum fit gives $v\sin i$=5$\pm$1\,\kms , $T_{\rm eff}$=5,670$\pm$140\,K, $\log g$=4.5$\pm$0.05, and [Fe/H]=--0.12$\pm$0.09 identical to within the error bars for both components. \Halpha\ appears never fully resolved in our spectra but the stronger photospheric lines are clearly doubled. The line ratio at $\lambda$672\,nm is 0.74$\pm$0.09 from  Ca\,{\sc i} 671.7\,nm, we designate the stronger-lined component with $a$ and the weaker with $b$. Both components show a Li\,{\sc i} 670.8-nm absorption line as strong as the adjacent Ca\,{\sc i} 671.7 line. For the $a$ component, we measure a Li\,{\sc i} 670.8-nm line with an equivalent width of 55$\pm$3\,m\AA\ (compared to 60$\pm$3\,m\AA\ for Ca\,{\sc i} 671.7). For the $b$ component the Li EW is 38$\pm$3\,m\AA\ (compared to 44$\pm$3\,m\AA\ for Ca\,{\sc i}). The Li lines appear blended with the close-by (weak) Fe\,{\sc i} 670.74-line. The latter amounts to $\approx$6\,m\AA \ in spectra of the Sun (sky) taken with the same equipment. We subtract this equivalent width from above values and multiply with the respective line ratio and obtain a logarithmic abundance of $\log n$=2.6$\pm$0.2 (85\,m\AA) for $a$ and $\log n$=2.5$\pm$0.2 (76\,m\AA) for $b$ (NLTE tables of Pavlenko \& Magazz\'u \cite{pm96}). We note that these values are more uncertain than the others in this paper because the effective temperature is from the composite spectrum.

A FFT of the CoRoT photometry has the highest peak at 12~d, while the ACF peaks at 25~d. We cannot discern which one is the alias but the measured $v\sin i$, combined with the assumption of a radius of a G3V star from $T_{\rm eff}$ and $\log g$, suggests a period of around 10~d.

\subsection*{\object{HSS 63} = \#32}

Not a cluster member. The mean RV from 6 measurements is --2.6$\pm$1.5\,\kms. The PARSES effective temperature is $\approx$6000\,K and $\log g\approx 3.8$. $v\sin i$ is $\approx$20\,\kms . Variable [O\,{\sc i}] emission lines at 557.7, 630.0, and 636.4~nm are present with line intensities relative to the continuum of 8--60, 1--4, and $<$1, respectively. The DIB line at $\lambda$661.3 appears very strong and has an EW of 130\,m\AA . Sodium D lines appear sharp and doubled. Therefore, the star is likely a background target.

\subsection*{\object{HSS 40} = \#33}

The mean RV from 16 measurements is --12.4$\pm$8\,\kms. Therefore, the stars is formally not a cluster member but the spread of the individual RVs is unusual high. The CoRoT photometry shows no sign for rotational modulation. \Halpha\ and \Hbeta\ have a shallow core and broad wings and possibly show an asymmetric profile. Variable [O\,{\sc i}] emission lines at 557.7, 630.0, and 636.4~nm are present with line intensities relative to the continuum of $\approx$10, 0.4--4.0, and 0.1--0.8, respectively. The interstellar Na\,{\sc i}\,D lines are very strong and reach basically zero intensity in the line core. The DIB line at $\lambda$661.3 appears also very strong and has an EW of 133\,m\AA . We conclude that the star is a background target.

\subsection*{\object{HSS\,74} = \#34}

The average radial velocity of --25.7\,\kms\ indicates cluster membership but its rms of 6\,\kms\ indicates it may also be a binary. The PARSES values are only marginally constrained and yield an effective temperature of 5500$\pm$400\,K, $\log g$=4.4$\pm$0.7, $v\sin i$=7$\pm$4\,\kms\ and [Fe/H]=--0.18$\pm$0.10. We detect a strong Li\,{\sc i} $\lambda$670.8-nm line with an equivalent width of 70$\pm$5~m\AA\ (compared to 102$\pm$5\,m\AA\ for the nearby Ca\,{\sc i} 671.7 line). The line is blended with the nearby Fe\,{\sc i} 670.74-line and we subtract 6\,m\AA\ from the measured EW based on the solar spectrum. This EW converts to a logarithmic Li abundance of $\log n$=2.2$\pm$0.1 (NLTE). \Halpha\ appears very asymmetric possibly with red-shifted emission.

The CoRoT data show a rotationally modulated light curve with a full amplitude of $\approx$0\fm025 and a period of 10.1\,d. Modulated onto this variation are numerous micro flares. We counted approximately 200 individual events within 78\,d with peak amplitudes on the same order of the rotational modulation. This is one of the most extreme flare stars we have ever seen.

\subsection*{\object{HSS 222} = \#35}

The best exposed of the five spectra show a Li\,{\sc i}  $\lambda$670.8-nm absorption line with an equivalent width of 72$\pm$7~m\AA\ (compared to 130$\pm$10\,m\AA\ for the nearby Ca\,{\sc i} 671.7 line). The spectrum synthesis allows only an estimate of $T_{\rm eff}$=5770$\pm$100\,K, $\log g$=4.52$\pm$0.15, [Fe/H]=--0.12$\pm$0.04 and $v\sin i$=5$\pm$1\,\kms\ but, if taken at face values, convert the de-blended equivalent width (66\,m\AA) to a logarithmic NLTE abundance of $\log n$=2.45$\pm$0.1 consistent with other cluster members. The photometric amplitude of $\approx$20\,mmag is the largest of all stars in our sample. The light curve shows a double-humped shape with a period of 9.85\,d that we interpret to be the rotation period of the star. With the $v\sin i$ of 5\,\kms\ the inclination of the rotational axis must be around $\approx$75\degr . 

\subsection*{\object{HSS 240} = \#36}

The spectrum synthesis suggests $T_{\rm eff}$=5760$\pm$70\,K, $\log g$=4.6$\pm$0.1, and a metallicity of --0.10$\pm$0.04. Best fit is with $v\sin i$=4$\pm$1\,\kms .  Lithium $\lambda$670.78\,nm is detected with an equivalent width of 90$\pm$10\,m\AA\ compared to the Ca\,{\sc i}~671.7 line with 102\,m\AA . Given the effective temperature of 5760\,K from the PARSES fit, the de-blended EW of 84\,m\AA\ converts to a logarithmic Li abundance of $\log n$=2.6$\pm$0.1 (NLTE). The main period from the FFT is at 11.4492\,d but is split by a second peak that is visible in the ACF. A formal conversion into $\Delta\Omega/\Omega$ would give 0.421$\pm$0.001. A smaller peak in the FFT at 4.92594\,d could be produced by a spot group separated by 180 longitudinal degrees from the main group and seen for only a short time. This star shows most likely a high degree of spot evolution instead of differential rotation, explaining the otherwise unusual large value for $\Delta\Omega/\Omega$ (see also target \#38).

\subsection*{\object{HSS 165} = \#37}

We measure a Li\,{\sc i} 670.8-nm line with an equivalent width of 83$\pm$3~m\AA\ (compared to 124$\pm$10\,m\AA\ for the nearby Ca\,{\sc i} 671.7 line). Because of the low $v\sin i$ of $<$2.5\,\kms\ the Li line is not blended with the nearby Fe\,{\sc i} 670.74-line and thus the effective temperature of 5790$\pm$70\,K, $\log g$=4.65$\pm$0.07 and [Fe/H]=--0.03$\pm$0.04 from STELLA/SES convert the equivalent width into a Li abundance of $\log n$=2.6$\pm$0.1 under NLTE.

\subsection*{\object{USNO-B1 0953-0376947} = \#38}

The FFT shows large frequency splitting around a period of 8.816\,d which would suggest a formal $\Delta\Omega/\Omega$ of 0.72$\pm$0.01, nearly four times the solar value. The ACF shows the modulation as well but due the low rotation rate of the star with respect to the length of the observation, the signal is more likely to be due to spot evolution rather than due to differential rotation.

We measure a Li\,{\sc i} $\lambda$670.8-nm line with an equivalent width of 80$\pm$8~m\AA\ (compared to 104$\pm$5\,m\AA\ for the nearby Ca\,{\sc i} 671.7 line). The effective temperature of 5500\,K, $\log g$=4.6 and [Fe/H]=--0.25$\pm$0.1 from PARSES suggests a mid-G dwarf consistent with the \Halpha\ morphology. $V\sin i$ is 7$\pm$2\,\kms . Assuming 6\,m\AA\ contribution from the Fe\,{\sc i} blend, we obtain a Li abundance of $\log n$=2.3$\pm$0.1 (NLTE).

%-------------------------------------------------------------------------
\end{document}